\newif\ifarxiv
\arxivtrue

\documentclass[runningheads]{llncs}
\usepackage[T1]{fontenc}
\usepackage{graphicx}
\usepackage{substr}
\usepackage{tikz}
\usepackage{collcell}

\usepackage{etoolbox}

\newtoggle{inTableHeader}%
\toggletrue{inTableHeader}%
\newcommand*{\StartTableHeader}{\global\toggletrue{inTableHeader}}%
\newcommand*{\EndTableHeader}{\global\togglefalse{inTableHeader}}%

\let\OldTabular\tabular%
\let\OldEndTabular\endtabular%
\renewenvironment{tabular}{\StartTableHeader\OldTabular}{\OldEndTabular\StartTableHeader}%

\newcommand*{\MinNumber}{0.0}%
\newcommand*{\MidNumber}{0.0001} %
\newcommand*{\MaxNumber}{100.0}%
\newcommand*{\complTableColWidth}{2em}

\newcommand{\ApplyGradient}[1]{%
  \iftoggle{inTableHeader}{#1}{
    \ifdim #1 pt > \MidNumber pt
        \pgfmathsetmacro{\PercentColor}{max(min(100.0*(#1 - \MidNumber)/(\MaxNumber-\MidNumber),100.0),0.00)} %
        \hspace{-0.33em}\colorbox{red!\PercentColor!white}{\parbox{\complTableColWidth}{\hfill #1}}
    \else
        \pgfmathsetmacro{\PercentColor}{max(min(100.0*(\MidNumber - #1)/(\MidNumber-\MinNumber),100.0),0.00)} %
        \hspace{-0.33em}\colorbox{white!\PercentColor!white}{\parbox{\complTableColWidth}{\hfill #1}}
    \fi
  }}

\newcolumntype{R}{>{\collectcell\ApplyGradient}r<{\endcollectcell}}

\usepackage{booktabs}   
\usepackage{colortbl}

\usepackage{amsmath}
\usepackage{mathtools}
\usepackage{stmaryrd}
\usepackage{xspace}

\newcommand*{\slstar}{\mathbin{\ast}} 
\newcommand*{\pointsto}{\mapsto} 

\usepackage{listings}

\newcommand*{\ass}{\coloneqq}
\newcommand*{\lstcodestyle}{\ttfamily\normalsize}

\makeatletter
\lst@InstallKeywords k{types}{typestyle}\lstcodestyle{typestyle}{}ld
\lst@InstallKeywords k{names}{namestyle}\lstcodestyle{namestyle}{}ld
\lst@InstallKeywords k{expressions}{expressionstyle}\lstcodestyle{expressionstyle}{}ld
\makeatother

\lstdefinelanguage{default}{
  basicstyle=\lstcodestyle,
  morekeywords={method,var,new,while,if,else,return,assume,assert,inhale,exhale,fold,unfold,package,apply},
  literate={~}{\ }1{:=}{$\ass$}1{!}{$\neg$}1{==}{$=$\ }1{!=}{$\neq$}1{<=}{$\leq$}1{>=}{$\geq$}1{&&}{$\slstar$\ }1{||}{$\vee$}1{==>}{$\implies$}1{**}{$\sep$}1{->}{$\mapsto$}1{===}{$\equiv$}1,
  morecomment=[l]{//},
  morecomment=[s]{/*}{*/},
}

\newcommand{\code}{\lstinline[language=default, mathescape=true]} 

\newcommand*{\ie}{i.e., }

\newcommand*{\eg}{e.g., }

\newcommand*{\figref}[1]{Fig.~\ref{#1}}
\newcommand*{\secref}[1]{Sec.~\ref{#1}}

\newcommand\finalVersion[1]{#1}
\newcommand\lastMinute[1]{#1}

\newcommand\malte[1]{#1}

\newcommand*{\greedy}{SE-PS\xspace}
\newcommand*{\mce}{SE-PC\xspace}
\newcommand*{\sica}{SE-TR\xspace}
\newcommand*{\sicax}{SE-TR}
\newcommand*{\caco}{VCG-TR\xspace}
\newcommand*{\cacox}{VCG-TR}
\newcommand*{\carbon}{VCG-TA\xspace}
\newcommand*{\carbonx}{VCG-TA}

\newcommand*{\gprustiother}{$\textit{Ru}_1$\xspace}
\newcommand*{\gprusticp}{$\textit{Ru}_2$\xspace}
\newcommand*{\gmendel}{$\textit{Ru}_3$\xspace}
\newcommand*{\ggobra}{\textit{Go}\xspace}
\newcommand*{\ggobracs}{$\textit{Go}_{C}$\xspace}
\newcommand*{\gdafnygobra}{$\textit{Da}_G$\xspace}
\newcommand*{\gdafnyviper}{$\textit{Da}_V$\xspace}
\newcommand*{\grsl}{\textit{RSL}\xspace}
\newcommand*{\gtwovyperNoRes}{\textit{SC}\xspace}
\newcommand*{\gtwovyperRes}{$\textit{SC}_R$\xspace}
\newcommand*{\gmpp}{\textit{PP}\xspace}
\newcommand*{\gnaginiNormal}{\textit{Py}\xspace}
\newcommand*{\gnaginimpp}{$\textit{Py}_P$\xspace}
\newcommand*{\greach}{\textit{Rea}\xspace}
\newcommand*{\gviper}{\textit{Vi}\xspace}
\newcommand*{\gvoila}{\textit{Vo}\xspace}
\newcommand*{\gvercorsCase}{$\textit{Ve}_{C}$\xspace}
\newcommand*{\gvercors}{$\textit{Ve}_{V}$\xspace}
\newcommand*{\ISCs}{quantified resources\xspace}

\newcolumntype{g}{>{\columncolor{light-gray}}c}

\usepackage{amssymb}

\usepackage{upgreek}

\renewcommand{\sigma}{\upsigma}

\newcommand*{\terminology}[1]{\textit{#1}}
\newcommand*{\abbr}[1]{\textbf{\underline{#1}}}

\newcommand*{\citeauthor}[2]{#1~\cite{#2}}

\newcommand*{\limplies}{\Rightarrow}
\newcommand*{\lthen}{\mathbin{?}}
\newcommand*{\lelse}{\mathbin{:}}

\begin{document}
\title{Verification Algorithms for Automated Separation Logic Verifiers}
\author{
Marco Eilers 
\and Malte Schwerhoff 
\and Peter M\"uller 
}
\authorrunning{M. Eilers et al.}
\institute{Department of Computer Science, ETH Zurich, Switzerland \\
\email{\{marco.eilers, malte.schwerhoff, peter.mueller\}@inf.ethz.ch}}
\maketitle              %
\begin{abstract}
Most automated program verifiers for separation logic use either symbolic execution or verification condition generation to extract proof obligations, which are then handed over to an SMT solver. Existing verification algorithms are designed to be sound, but differ in performance and completeness. These characteristics may also depend on the programs and properties to be verified. Consequently, developers and users of program verifiers have to select a verification algorithm carefully for their application domain. Taking an informed decision requires a systematic comparison of the performance and completeness characteristics of the verification algorithms used by modern separation logic verifiers, but such a comparison does not exist.

This paper describes five verification algorithms for separation logic, three that are used in existing tools and two novel algorithms that  combine characteristics of existing symbolic execution and verification condition generation algorithms. A detailed evaluation of implementations of these five algorithms in the Viper infrastructure assesses their performance and completeness for different classes of input programs. Based on the experimental results, we identify candidate portfolios of algorithms that maximize completeness and performance.

\keywords{Symbolic execution, verification condition generation, separation logic, heap representation, SMT solver, portfolio}
\end{abstract}
\section{Introduction}
\label{sec:intro}

Given a program and a specification, automated deductive program verifiers such as Boogie~\cite{LeinoBoogie2}, Corral~\cite{LalQ14}, Dafny~\cite{Leino10}, and Why3~\cite{FilliatreP13} compute \emph{proof obligations} whose validity implies the correctness of the input program. These proof obligations are typically checked using SMT solvers, such as CVC5~\cite{CVC5} or Z3~\cite{Z3}.

For program verifiers based on separation logic~\cite{Reynolds02a} or related permission logics~\cite{SmansJP09}, proof obligations are computed using two prevalent verification algorithms: symbolic execution (SE) and verification condition generation (VCG). For instance, Caper~\cite{Dinsdale-YoungP17}, Gillian~\cite{MaksimovicASG20}, JaVerT~\cite{SantosMSG19}, SecC~\cite{ErnstM19}, Smallfoot~\cite{BerdineCO05}, and VeriFast~\cite{JacobsSPVPP11} are separation logic verifiers based on symbolic execution, whereas Chalice~\cite{LeinoMS09} and GrassHopper~\cite{PiskacWZ14} use verification condition generation. Viper~\cite{viper} provides two backend-verifiers, one based on SE and one on VCG\@.

Even though these tools differ in many aspects of the supported programming language, separation logic, and proof automation, they employ fairly uniform SE and VCG algorithms. Their SE algorithms use a symbolic heap representation based on separation logic's \emph{partial-heap semantics}~\cite{ParkinsonS11}: a symbolic heap maps those separation logic resources (in particular, heap locations) to symbolic values that are owned in a given program state. Each owned resource is represented by one or more \emph{heap chunks}, which map resources to ownership and value information.
In contrast, the VCG algorithms implemented in separation logic verifiers use a \emph{total-heap representation}, in which the heap is a total map from memory locations to values, and the currently-owned resources are tracked in a separate data structure. 
These two different ways of \emph{internally} modeling the heap both implement the same \emph{source-level} language semantics.

These verification algorithms, and their variations implemented in various tools, are all designed to be sound, but strike different trade-offs between performance and completeness. For instance, SE verifies each path through a method separately, whereas VCG typically generates one proof obligation for the entire method. Therefore, VCG produces fewer, but larger proof obligations, which may affect the effectiveness and performance of the underlying SMT solver.

Consequently, developers of program verifiers need to choose the verification algorithm carefully, depending on the intended application area of their tools. For verifiers that support several algorithms, such as the verifiers built on top of the Viper infrastructure~\cite{vercors,nagini,gobra}, this choice needs to be made by users. Taking an informed decision requires a systematic comparison of the performance and completeness characteristics of the verification algorithms used by modern separation logic verifiers. 
Such a comparison necessitates implementations of all relevant algorithms for the same programming language, verification logic, and tool because comparisons across different settings would not yield meaningful results. To the best of our knowledge, such implementations and, consequently, a comprehensive comparison do not exist.

\subsubsection{This Work.} This paper describes the following five verification algorithms and performs a detailed comparison. 

\begin{enumerate}
\item \textbf{\greedy:} 
An SE algorithm that looks up information in the \emph{partial} symbolic heap by trying to identify a \emph{single} heap chunk to provide the required information. This algorithm is used in 
JaVerT, SecC, VeriFast, and Viper's SE-backend.

\item \finalVersion{\textbf{\mce:}} 
An SE \emph{partial} heap algorithm that performs look-ups by \finalVersion{\emph{combining}} the information available in all heap chunks. Combining different heap chunks may provide additional information, for instance, by summing up fractional permissions~\cite{Boyland03} or by using disjunctive properties. 

\item \textbf{\sica:} 
An SE algorithm that uses a \emph{total} heap representation per individual \emph{resource}, akin to \caco below. 

\item \textbf{\caco:} 
A VCG \emph{total} heap algorithm that uses a separate map per \emph{resource}. This representation is used by GrassHopper.

\item \textbf{\carbon:} 
A VCG \emph{total} heap algorithm that stores the information for \emph{all} resources in a single map. This representation is used in Chalice and Viper's VCG-backend.
\end{enumerate}

\mce and \sica are novel algorithms, which introduce characteristics of existing VCG algorithms into SE, namely simultaneous reasoning about multiple chunks of the (partial) heap, and a total heap representation, respectively. Thereby, they offer different trade-offs than existing algorithms. 

To enable a fair comparison, we implemented all five algorithms for Viper. We evaluated them on a diverse benchmark suite that includes existing Viper examples and code produced by different Viper frontends, which allows us to draw conclusions for different kinds of input programs.

Our comparison identifies \greedy as the best algorithm overall, but shows that the different verification algorithms have complementary strengths. Based on our findings, we identify and discuss several \emph{portfolios} of algorithms, which maximize completeness across the benchmark suite. In deductive verification, portfolio approaches have been used successfully for the underlying SMT solver~\cite{FilliatreP13,portfolio24}, but, to our knowledge, not for the equally-important verification algorithms.

\subsubsection{Contributions and Outline.} 

We make the following contributions:

\begin{itemize}
\item We survey the SE and VCG algorithms used in existing separation logic verifiers and propose two new algorithms, which combine characteristics of existing SE and VCG algorithms (\secref{sec:algorithms}).

\item We provide the first systematic comparison of verification algorithms for separation logic. A diverse set of benchmarks provides insights into the performance and completeness for different classes of input programs (\secref{sec:evaluation}).

\item We identify candidate portfolios of verification algorithms to maximize performance and completeness, several of which include \sica, one of the novel algorithms we propose (\secref{sec:portfolio}).
\end{itemize}

\noindent

\section{Verification Algorithms}
\label{sec:algorithms}

In this section, after providing necessary background on the Viper language, we discuss the two main design dimensions for verification algorithms for separation logic (SE vs.\ VCG, and total vs.\ partial heap representations), give an overview of the considered algorithms, and discuss various design trade-offs.

\subsection{Viper Verification Language}
\label{sec:viper_verification_language}

The Viper language~\cite{viper} is a simple object-based imperative language with specification features like pre- and postconditions and loop invariants. 
Viper is based on implicit dynamic frames~\cite{SmansJP09}, a variant of separation logic, and supports advanced separation logic features such as \terminology{fractional permissions}~\cite{Boyland03}, \terminology{predicates}~\cite{ParkinsonB05}, \terminology{magic wands}, and \terminology{quantified resources}~\cite{0001SS16} (also called iterated separating conjunctions). 
Verification algorithms for Viper have to support all of these features, which makes Viper an interesting target for a comparison.

A Viper state consists of local variables and a built-in heap that maps locations (consisting of a reference and a field) to values. Control flow is expressed via conditionals, loops, method calls, and gotos. Whereas statements may have side effects, expressions are always side-effect free and include calls to (partial) functions, which may inspect the heap.

Following the implicit dynamic frames approach, Viper assertions express resource ownership separately from value information. For example, the assertion \code{acc(x.f) && x.f == 1} (corresponding to separation logic's points-to predicate \code{x.f $\pointsto$ 1}) includes an \terminology{accessibility predicate} \code{acc(x.f)}, expressing exclusive ownership of the heap location, and a \terminology{heap-dependent expression} to constrain its value.
The general shape of accessibility predicates is \code{acc($R$, $p$)}, where $R$ denotes a \terminology{resource} and $p$ a fractional permission. Resources can be heap locations, predicate instances, and magic wands; all resources can be universally quantified over. Predicates abstract over (possibly unbounded) heap data structures, whereas magic wands are used to express partial data structure, which occur, for instance, during iterative traversals. 

\subsection{Design Dimensions}
\label{sec:overview_verifier_implementations}

Verification algorithms for separation logic can be classified according to the technique they use to compute proof obligations (SE or VCG) and according to their heap representation (total or partial). In the following, we survey these dimensions and their main trade-offs.

\subsubsection{SE vs.\ VCG\@.}
\label{sec:se_vs_vcg}

Verification in separation logic is modular, that is, each method is verified independently, using method specifications to reason about calls. SE and VCG differ in how they compute the proof obligations for each method.

SE uses a symbolic state, typically a triple of symbolic store, heap, and path conditions. It explores each path through a method body separately (using loop invariants to represent a statically-unknown number of loop iterations). Statements on the path may update the symbolic state; in particular, the conditions of if-statements and loops are recorded in the path conditions. Expressions and assertions are evaluated in the symbolic state. Proof obligations, for instance, to show that an assertion holds, are expressed over the current symbolic state and discharged on the fly via an SMT query. Consequently, SE typically generates many SMT queries for each method body.

In contrast, VCG uses a predicate transformer, usually weakest preconditions, to produce (typically) a single proof obligation (and, thus, SMT query) per method body. This predicate transformer is based on a state model that, in the context of separation logic, must encode heap and ownership information (\eg via a map axiomatization, see below).

There are two fundamental differences between SE and VCG. First, SE generates many, but comparably small and simple SMT queries, whereas VCG produces a single, more complicated query. This difference may affect verification times. Moreover, the complexity of the SMT queries can affect the SMT solver's ability to discharge (valid) queries.
Second, for a rich verification logic such as Viper's, an SE algorithm is complex and performs substantial work for maintaining the symbolic state, whereas VCG delegates most of the heavy lifting to the SMT solver. This makes SE more difficult to implement, but also offers the potential for many optimizations (possibly with the use of additional SMT queries), whereas it is more difficult to direct the proof search of an SMT solver. Our evaluation in the next section explores these trade-offs and others.

\subsubsection{Partial vs.\ Total Heaps.}
\label{sec:partial_vs_total_heaps}

Both SE and VCG need to represent heap and ownership information. Existing SE algorithms do that by maintaining an \emph{internal} map data structure, typically a \emph{collection of heap chunks} (also called \terminology{heaplets}). A heap chunk is typically a tuple $(x,f,v,p)$, denoting $p$ permission to memory \finalVersion{location $x.f$ at which value $v$ is stored}. Resources that are not owned in a state have no corresponding heap chunk (or a chunk with permission amount $p=0$). Consequently, these internal map data structures represent \terminology{partial heaps}, which represent value and ownership information simultaneously. 

In contrast, VCG algorithms use an \emph{external} representation that tracks heap information only as part of the SMT queries. Since maps in SMT are total, value and ownership need to be encoded separately as two total maps $H$ and $M$. The \terminology{heap} $H: \mathit{Resource} \mapsto \mathit{Value}$ maps resources to their values, whereas the 
\terminology{permission mask} $M: \mathit{Resource} \mapsto \mathit{Permission}$ tracks ownership by mapping each resource to the  permission amount currently held ($1$ for exclusive ownership, and $0$ if the resource is not owned in the current state). Suitable proof obligations ensure that $H$ is accessed only at resources for which $M$ contains the necessary permission. That is, the mask effectively represents the domain of a partial heap. We call this representation \terminology{total heaps}. 

We highlight three key differences between partial and total heaps here. 
First, total heaps generally lead to more complex SMT queries. In particular, each change of the heap (or mask) leads to an SMT term that relates the new heap to the previous one, leading to increasingly large formulas, whereas the data structure for partial heaps can be updated destructively. Moreover, encoding the heap information in an SMT query typically uses many universal quantifiers for total heaps, whereas partial heaps are finite collections whose content can be described in quantifier-free formulas.

Second, partial heaps generally require more complex algorithms to perform heap look-ups and modifications, possibly involving SMT queries. In contrast, total heaps delegate much of the heavy lifting to the SMT solver. This difference is especially prominent for resources that represent an unbounded number of heap locations, such as recursive predicates and iterated separating conjunction. These require dedicated data structures and operations in partial heaps~\cite{0001SS16}, but fairly trivial encodings with total heaps.

Third, total heaps greatly simplify the encoding of heap-dependent functions to SMT, as uninterpreted functions of the total heap and corresponding axioms~\cite{HeuleKMS13}. In contrast, partial heaps require non-trivial algorithms to extract the information needed to determine a function's value~\cite{Schwerhoff16}.

\subsection{Algorithms}
\label{sec:five-algorithms}

In this subsection, we sketch five verification algorithms that occupy different spots in the design space described above and, thus, have different performance and completeness characteristics. 
\finalVersion{
Note that these algorithms do not directly correspond to the four combinations of heap representation and technique used to compute proof obligations explained before:
First, there is no algorithm combining VCG and partial heaps, since VCG algorithms necessarily require an external heap representation.
Second, we discuss two different algorithms that combine SE with partial heaps. 
}
Three of \finalVersion{the} algorithms are used in existing tools, whereas two are novel SE algorithms, including the first SE algorithm that uses total heaps. We will see later in \secref{sec:portfolio} that these new algorithms complement existing ones, which makes them especially useful for portfolio approaches.

We focus the following presentation on two core operations: evaluating an expression, as well as consuming an assertion, which includes checking that it holds and removing its resources from the current state.

\subsubsection{\greedy.}
This algorithm combines symbolic execution with the \abbr{p}artial heap model and is used by the existing SE tools for separation logic. Evaluating a source-level heap read $x.f$ is performed by trying to find a chunk $(y,f,v,p)$ in the symbolic heap such that $x = y$ and $0 < p$. If such a chunk can be found, the result of the symbolic evaluation is $v$. Otherwise, verification fails.
Analogously, consuming $q$ permissions to a heap location $x.f$ is implemented by finding a chunk $(y,f,v,p)$ such that $x = y$ and $q \leq p$. If found, the chunk is replaced by $(y,f,v,p-q)$; otherwise, verification fails.
Finding matching chunks in general requires SMT queries to account for aliasing; in practice, however, syntactical checks often suffice, and can significantly reduce the number of SMT queries.

Note that both operations are performed on a \abbr{s}ingle heap chunk, which may lead to incomplete heap information and, thus, spurious errors. For instance, when the permission to a heap location is split over several chunks, this algorithm will use only one of them (rather than computing the total sum of permission amounts) and might, thus, report a verification error if the permission amount in that one chunk is not sufficient to perform an operation. To reduce the number of such spurious errors, the algorithm performs various \emph{state consolidation} steps at heuristically determined points (\eg triggered by an imminent verification failure). For instance, it may merge two heap chunks (\ie add their permission amounts) if the SMT solver can prove that they refer to the same resource. \finalVersion{State consolidation may also introduce non-aliasing constraints, \ie assume for any pair of chunks $(x_1,f,v_1,p_1)$ and $(x_2,f,v_2,p_2)$ that $x_1 \neq x_2$ if $p_1 + p_2 > 1$.}

State consolidation eliminates some spurious errors, but performing operations on a single heap chunk remains incomplete, for instance, in situations with \terminology{disjunctive aliasing}. In a state where  $x = y \lor x = z$ and where permissions to both $y.f$ and $z.f$ are available, consuming permission to a location $x.f$ fails because the algorithm cannot find a \emph{single} chunk that definitely provides the necessary permission. To work around this issue, users can force the SE to branch on the disjunction (\eg by inserting if-statements), such that a single chunk can be found on each branch.

\subsubsection{\mce.}
To address the shortcomings of \greedy, we designed a novel variation that also uses \abbr{p}artial heaps, but consults \finalVersion{and \abbr{c}ombines} information from all chunks.
Evaluating a source-level heap read $x.f$ summarizes facts scattered across all relevant heap chunks $(y_1,f,v_1,p_1), \ldots, (y_n,f,v_n,p_n)$: The effective value of $x.f$ is denoted by a fresh symbol $v$ that is defined by the new path condition $(x = y_1 \limplies v = v_1) \land \ldots \land (x = y_n \limplies v = v_n)$. Analogously, the effective permission to $x.f$ is denoted by a fresh symbol $p$ that is defined by the symbolic expression $p = (x = y_1 \lthen p_1 \lelse 0) + \ldots + (x = y_n \lthen p_n \lelse 0)$. 
Consuming $q$ permission to a heap location $x.f$ similarly may remove fractions of $q$ from different heap chunks.

Compared to \greedy, \mce effectively shifts work from the SE algorithm to the SMT solver: it reduces the number of state consolidation steps (but does not entirely eliminate them), at the price of more complex path conditions and SMT queries. The next algorithm pushes this trade-off even further.

\subsubsection{\sica.}
Even though all existing SE algorithms for separation logic use partial heaps, SE is also compatible with \abbr{t}otal heaps, as this novel algorithm shows. It uses a heap/mask pair $(H_R, M_R)$ for each kind of \abbr{r}esource (\ie field or predicate) $R$.
Evaluating a source-level heap read $x.f$ simply asserts $0 < M_f[x]$ and produces the symbolic look-up expression $H_f[x]$, where $H_f$ and $M_f$ are the current heap and mask component of the symbolic state.
Consuming $q$ permission checks $q \leq M_f[x]$ and then replaces the symbolic state's mask $M_f$ with an updated version $M'_f = M_f[x \mapsto M_f[x] - q]$. The necessary map update axioms are part of the heap's background axiomatization that is given to the SMT solver.
To prevent the verifier from unsoundly framing information about heap locations for which no permission is held \finalVersion{and which may thus be modified by whoever has obtained the permission}, they are assigned non-deterministic values.

Using a total heap eliminates the need for state consolidation because all information about a resources is represented by a single heap/mask pair, rather than multiple chunks. \finalVersion{Non-aliasing can be assumed using a global axiom stating that for all masks $M_R$ and receivers $x$, $M_R[x] \leq 1$.} Nevertheless, the algorithm retains some of the key benefits of SE, such as cheap syntactical comparisons, which are sufficient in many cases. However, compared to partial heap algorithms, it complicates SMT queries, which now require a theory for maps or a suitable axiomatization.

\subsubsection{\caco.}
VCG algorithms do not have an internal representation of the heap and, therefore, they necessarily use \abbr{t}otal heaps, which can be encoded in SMT\@. One option, implemented in GrassHopper, is to use a heap/mask pair per \abbr{r}esource, like in the previous algorithm.
Evaluating source-level heap reads, or consuming permissions, are incorporated into the verification condition as described for the previous algorithm. Doing that in a VCG algorithm leads to the advantages and disadvantages outlined in \secref{sec:overview_verifier_implementations}.

\subsubsection{\carbon.}
A variation of \caco that uses a single heap/mask pair across \abbr{a}ll resources. It is used, for instance, in VeriCool~\cite{SmansJP09} and Viper's VCG-backend.
Heap reads and mask updates are encoded as described for \sica, with the only change that the field becomes another index into the single heap or mask.

Using a single heap simplifies, 
for example, the encoding of predicates (see \secref{sec:viper_verification_language}) but, 
on the other hand, complicates framing for heap-dependent functions, since updating \emph{any} resource changes the (only) heap and, thus, requires proof steps to show that other resources are not affected.

\bigskip
The discussion of these five algorithms illustrates various design choices, which may affect performance, for instance, by shifting work between the verification algorithm and the SMT solver. These choices also affect completeness. Most prominently, algorithms using total heaps make heavy use of universal quantifiers, making the SMT queries undecidable. In practice, verification tools use the SMT solver's E-matching~\cite{ematching}, which allows them to guide quantifier instantiations by specifying matching patterns (also called triggers). However, making those too strict can prevent necessary instantiations (causing spurious errors), whereas making them too permissive may cause too many unnecessary instantiations (and, thus, bad performance).
It is thus crucial to assess performance \emph{and} completeness of verification algorithms empirically, as we do next.

\section{Evaluation}
\label{sec:evaluation}

This section presents our empirical evaluation. We first discuss relevant implementation details of the algorithms, introduce the benchmarks, and describe our set-up. We then present and interpret our evaluation results, in terms of completeness and performance of the different algorithms. Finally, we conclude by discussing potential threats to the validity of our results.

\subsection{Implementations}

Viper's two existing backends implement \greedy and \carbon, respectively. 
We have extended Viper to implement the remaining three algorithms (for the full Viper language as of version 23.07), reusing parts of the existing implementations where possible:
We based \mce on Viper's \greedy backend, which allowed us to reuse the entire SE engine and the state representation, but required re-implementing all heap-related parts of the algorithm.
We also based \sica on Viper's \greedy backend. Here, we could still re-use the SE engine, but had to re-implement the heap representation, all heap-manipulating operations, and code for axiomatizing heap-dependent functions.
Lastly, we based \caco on Viper's \carbon backend, which encodes a Viper program into a Boogie~\cite{boogie} program; Boogie then computes a verification condition and interacts with Z3. We reused this entire mechanism and the general encoding of statements and expressions, but had to adapt all heap-handling code.
All SE algorithms implement various \finalVersion{optimizations: %
they perform syntactic equality checks (\eg in \greedy to find a matching heap chunk) to avoid SMT queries, they simplify terms on the fly (\eg \sica, to keep mask terms simple, simplifies a mask $M_R$ to which the same permission has been added and later removed to be just $M_R$ again), they cache values (\eg the term resulting from reading a field in \mce), they actively query the SMT solver to check if paths can be pruned, and they optimize their communication with the SMT solver to avoid repeating large terms. 
} 
Some of \finalVersion{these optimizations} are crucial for scaling the algorithms to large examples, while others are only relevant for corner cases.
In the VCG algorithms, most of these optimizations are not possible; thus, they mainly rely on Boogie to generate efficient verification conditions~\cite{BarnettL05}.

\subsection{Benchmark Selection}

For our comparison, we selected a total of 537 example programs to be verified.
About 80\% of these were generated by one of several Viper frontends from programs written in different source languages, the other programs were manually written in Viper.
Each example represents a meaningful verification task (\eg from a publication's case study or verification competition): in particular, we excluded programs that represent regression tests or that test specific features in isolation.
The majority of the examples (388) is expected to verify; for the remainder, the expected result is some set of verification failures.

The examples vary along several dimensions: 
source level language (\eg Rust, Java, Python),
frontend verifier (\eg we obtained Viper programs from different Rust verifiers),
application area and complexity (ranging from individual functions to large case studies, \eg to verify cryptographic security of network protocols),
verified properties (\eg memory safety, complex functional specifications, or hyperproperties such as secure information flow),
Viper features used to encode source languages and properties,
and code size (ranging from 15 to 99,110 lines of Viper code, with a mean/median of 2400/495 LOC).

To observe the effect that these variations have on the completeness or performance of different verification algorithms, we partitioned the examples into different groups, listed below. We first grouped by frontend verifiers, which we then further refined: \eg by application area or typical usage patterns of Viper features.
This resulted in the following groups:

\begin{itemize}
  \item \gprustiother represents Rust programs verified using Prusti~\cite{prusti}, which heavily use predicates and magic wands but no \ISCs.
  \item \gprusticp and \gmendel contain \finalVersion{unsafe Rust code encoded by prototype versions of two different Viper frontends~\cite{vytautas-thesis,mendel-paper}}; the former heavily mutates the heap, whereas the latter does not use Viper's heap at all.
  \item \ggobra contains smaller Go programs encoded by Gobra~\cite{gobra}. \ggobracs contains larger examples from two case studies that prove correctness and security of real-world implementations of security protocols~\cite{wireguard,verifiedscion}. 
  \item \grsl contains weak-memory programs generated by a frontend~\cite{rsl2viper} for Relaxed Separation Logic~\cite{RSL}.
  \item \gtwovyperNoRes and \gtwovyperRes contain smart contracts encoded to Viper by 2vyper~\cite{2vyper}; the former group does not use Viper's heap at all, whereas the latter uses \ISCs, and additionally generates a lot of branches.
  \item \gmpp contains a product program encoding~\cite{mpp} that is used to prove a 3-safety hyperproperty. The generated programs lightly use the heap and heavily utilize branching.
  \item \gnaginiNormal contains Python programs encoded by Nagini~\cite{nagini,marco-thesis}, including two case studies~\cite{igloo,forster} that prove complex functional properties. \gnaginimpp contains Nagini-generated programs that additionally use a product program encoding to prove information flow security of the original Python programs~\cite{nagini-mpp}.
  \item \greach proves reachability properties about graph-manipulating programs~\cite{reachability}; it heavily relies on heap-dependent functions and \ISCs.
  \item \gviper contains various programs directly written in Viper, including examples from publications on \ISCs~\cite{0001SS16} and magic wands~\cite{wands}, and from the VerifyThis verification challenge~\cite{verifythis}.
  \item \gvercorsCase contains examples encoded by VerCors~\cite{vercors} that stem from larger case studies \malte{that} verify properties of Java~\cite{vercors-redblack} and CUDA~\cite{vercors-prefixsum} programs as well as examples written in VerCors's custom (Java-like) PVL language~\cite{vercors-permutation}. \gvercors contains VerCors-encoded examples from VerifyThis.

  \item \gdafnyviper and \gdafnygobra contain Viper~\cite{frei} and Gobra versions~\cite{egli}, respectively, of examples from a verification textbook~\cite{Leino10,leino2023program}.
  \item \gvoila contains examples that heavily manipulate the heap, generated by Voila~\cite{voila}, a frontend automating the fine-grained concurrency logic TaDA~\cite{tada}.
\end{itemize}

In all groups, the examples combine several Viper features to encode the desired language semantics and properties; some groups use certain features very prominently, while others are more heterogeneous.
Overall, we believe that our examples and groups form a good representation of the many different usages of Viper, and that our results can be transferred to other separation logic verifiers.

\subsection{Experimental Setup}
\label{sec:setup}

We evaluated completeness and performance of the different algorithms for each of the examples.
Our test system uses an AMD Ryzen 5900X with 32GB of RAM running Ubuntu 23.10.
For the VCG algorithms, we use Boogie 2.15.9. Our SMT solver is Z3 4.8.7\footnote{While this is an older version of Z3, it is the default version used with Viper 23.07.}.
All algorithms run with the same Z3 options and the same axiomatization for background theories other than the heap (\eg  sequences).

We ran each algorithm five times on each example on a warmed-up JVM, each time with a timeout of ten minutes.
We let each algorithm  report all errors for every example (\ie we did not stop after some number of errors were found).
To ensure that we measure the total workload, we disabled all parallelization along the tool chains.
To account for the heuristics-driven nature of SMT solvers, we consistently varied Z3's random seeds: we picked five fixed seeds and always used the $i$th seed for the $i$th run of every algorithm-example combination.
Lastly, we measured the verification time of each algorithm on an empty Viper program to obtain their fixed startup overhead (\eg from starting Boogie), 
and subtracted, from all times of a given algorithm, the difference between its own overhead and the overall lowest overhead. In practice, this difference was at most 320ms.

\subsection{Completeness Results}

\begin{table}[t]
\scriptsize
\centering
\begin{tabular}{l|l|R|R|R|R|R|R|R|R|R|R}
&$\Sigma$&\multicolumn{2}{c|}{SE-PS}&\multicolumn{2}{c|}{SE-PC}&\multicolumn{2}{c|}{SE-TR}&\multicolumn{2}{c|}{VCG-TR}&\multicolumn{2}{c}{VCG-TA}  \EndTableHeader \\ \hline
All&537&5.4&3.7&5.4&3.7&7.4&3.7&8.8&3.7&13.8&7.8\\ \hline
 \gprustiother&156&0.0&0.0&0.6&0.6&3.8&0.0&9.0&0.6&18.6&5.1\\ \hline
 \gprusticp&11&0.0&0.0&0.0&0.0&45.5&45.5&54.5&54.5&63.6&63.6\\ \hline
 \ggobra&11&18.2&9.1&0.0&0.0&18.2&0.0&0.0&0.0&0.0&0.0\\ \hline
 \ggobracs&17&5.9&0.0&0.0&0.0&17.6&17.6&29.4&11.8&35.3&29.4\\ \hline
 \grsl&21&19.0&14.3&19.0&14.3&0.0&0.0&38.1&28.6&33.3&28.6\\ \hline
 \gtwovyperNoRes&8&0.0&0.0&0.0&0.0&0.0&0.0&0.0&0.0&0.0&0.0\\ \hline
 \gtwovyperRes&13&0.0&0.0&0.0&0.0&0.0&0.0&0.0&0.0&0.0&0.0\\ \hline
 \gmpp&38&0.0&0.0&0.0&0.0&0.0&0.0&0.0&0.0&0.0&0.0\\ \hline
 \gnaginiNormal&23&8.7&8.7&13.0&4.3&4.3&4.3&13.0&0.0&21.7&8.7\\ \hline
 \gnaginimpp&18&16.7&11.1&11.1&5.6&11.1&0.0&5.6&5.6&5.6&5.6\\ \hline
 \greach&16&93.8&62.5&93.8&62.5&37.5&37.5&12.5&0.0&18.8&6.3\\ \hline
 \gviper&46&2.2&2.2&2.2&2.2&4.3&0.0&4.3&4.3&8.7&6.5\\ \hline
 \gvercorsCase&19&0.0&0.0&10.5&10.5&42.1&5.3&10.5&5.3&10.5&10.5\\ \hline
 \gvercors&5&0.0&0.0&0.0&0.0&20.0&0.0&20.0&0.0&0.0&0.0\\ \hline
 \gdafnygobra&8&0.0&0.0&0.0&0.0&0.0&0.0&25.0&12.5&25.0&12.5\\ \hline
 \gdafnyviper&41&0.0&0.0&0.0&0.0&0.0&0.0&0.0&0.0&0.0&0.0\\ \hline
 \gvoila&34&2.9&2.9&2.9&2.9&11.8&11.8&0.0&0.0&20.6&17.6\\ \hline
 \gmendel&52&0.0&0.0&0.0&0.0&0.0&0.0&1.9&0.0&1.9&0.0\\
\end{tabular}
\vspace{1mm}
\caption{Incompletenesses per algorithm per example group. $\Sigma$ is the total across all groups. For each algorithm, we show first \finalVersion{the percentage of examples where it was incomplete for any reason, and then the percentage of examples where it was incomplete due to timeouts and inconsistent results.}}\label{tbl:completeness}
\end{table}

Table~\ref{tbl:completeness} shows the number of incompletenesses per algorithm and example group, 
\ie the number of examples for which the algorithm
reported unexpected errors, timed out, or reported inconsistent results over the five runs. The latter is typically caused by differences in the SMT solver's proof search, \eg due to different random seeds.

\subsubsection{Overall Results.} Every algorithm is able to report the desired result for over 86\% of the examples. 
However, there is a clear distinction: \carbon has the most incompletenesses with 13.8\% (and the most timeouts by far), followed by \caco with 8.8\% and \sica with 7.4\%. The two partial heap algorithms perform the best with 5.4\% \finalVersion{each}.

Thus, our first, perhaps surprising observation is that \finalVersion{for our test set,} (sufficiently optimized) \textbf{partial heap algorithms are more complete than total heap algorithms}. That is, performing heap reasoning in the verification algorithm is more effective than leaving it entirely to the SMT solver.
This conclusion is supported by further observations:
(1)~SE algorithms, with their greater potential for optimizations, generally outperform the VCG algorithms;
(2)~\greedy, despite its conceptual incompleteness, performs \finalVersion{identically to \mce overall}, which produces more complex SMT queries by summarizing heap chunks;
(3)~\carbon, with its single heap, has a much higher number of timeouts than \caco with its separate heaps per resource.

\subsubsection{Impact of Optimizations.} To evaluate how much \greedy's completeness depends on optimizations, we re-ran all examples with a version of \greedy with the majority of its optimizations disabled. The resulting algorithm performs notably worse, with 59 (instead of 29) incompletenesses. We thus conclude that \textbf{the good performance of \greedy is due to significant optimization efforts}.%

\subsubsection{Complementarity.} A pairwise comparison shows that, for each algorithm, there are a number of examples that this algorithm is incomplete on, but another algorithm is not. For example, while \finalVersion{\mce and \greedy perform identically in overall numbers, \mce is complete on three examples where \greedy is not and vice versa}.
Other pairs differ more strongly, with \greedy and \carbon forming the extreme pair: there are in total 75 examples for which only one of the two algorithms is complete. We thus conclude that \textbf{being able to use more than one algorithm is advantageous in practice}, and we explore this further in \secref{sec:portfolio}.

\subsubsection{Differences Between Groups.} 
\finalVersion{While comparisons of the different algorithms for our overall test set can give us an indication of their overall performance, their results may be skewed due to over- or underrepresentation of different patterns in our test set. Thus, a more important observation is that the}
number of incompletenesses per algorithm differs significantly between example groups.
The two algorithms using partial heaps  (\greedy, \mce) are both incomplete on 15 out of 16 examples in the \greach group, making them essentially unusable for this group. This indicates that the \textbf{heavy use of heap-dependent functions framed by \ISCs} is problematic for partial heap algorithms.
All total heap algorithms perform better: \sica is incomplete on 6 out of 16 examples, the VCG algorithms on~2 and~3.

The opposite is the case for group \gprusticp, where all total heap algorithms time out for at least 5 out of 11 examples, while both partial heap algorithms are complete for all examples. Other groups that also heavily manipulate the heap are \gprustiother and \gvoila; \finalVersion{both exhibit the similar tendencies}, and likewise for \ggobracs with its large and complex case studies. These observations suggest that \textbf{total heap algorithms \finalVersion{can} struggle with a large number of heap updates}. Correspondingly, the heap representation does \emph{not} affect completeness for the two groups that do not use the heap at all (\gmendel and \gtwovyperNoRes).

\subsubsection{Conclusions.}
Given the previous observations and interpretations, we can draw three final conclusions regarding completeness:
First, \textbf{the heap representation has a bigger impact on completeness than the verification mechanism} (SE or VCG), since all groups that have large differences in completeness numbers have the largest difference between the partial heap and the total heap algorithms.
Second, \textbf{certain example groups effectively require specific combinations} (\eg VCG + total for \greach, SE + partial for \gprusticp).
Third, a general-purpose separation logic verifier should \textbf{implement at least two algorithms to be reasonably complete}.

\subsection{Performance Results}

To compare the performance of different verification algorithms, we measured the run time of each of the five algorithms on each of the examples five times, as explained in \secref{sec:setup}. We discarded the shortest and longest run times and computed the mean of the remaining three, leading to one data point for each algorithm-example combination.

\subsubsection{Comparison Method.}
Comparing algorithms on an example is sensible only if each algorithm reports the same verification result (otherwise, an algorithm that always immediately fails would be the fastest). Therefore, we compare \emph{pairs} of algorithms on examples for which both report the same result; using pairs instead of all five algorithms minimizes the number of examples to discard.

For each pair of algorithms and example, we compute the \emph{relative percentage difference} (RPD) of the two mean run times $t_1$ and $t_2$, defined as $(t_2 - t_1) / (0.5 \cdot (t_2 + t_1)) \cdot 100$,
which relates the run time \emph{difference} to the  \emph{average} run time of the example. Consequently, RPDs are independent of the absolute run times, which allows us to compare algorithms across examples with vastly different run times.
An RPD of 0 means equal performance, a positive value means that the first algorithm was faster, with higher values indicating bigger differences: \eg $+66.6$ indicates that the first algorithm took half the time of the second. The maximum RPD is $+200$, obtained when the first algorithm is essentially instant compared to the second. Conversely, negative values mean that the second algorithm was faster.

\begin{figure}
\includegraphics[width=\textwidth]{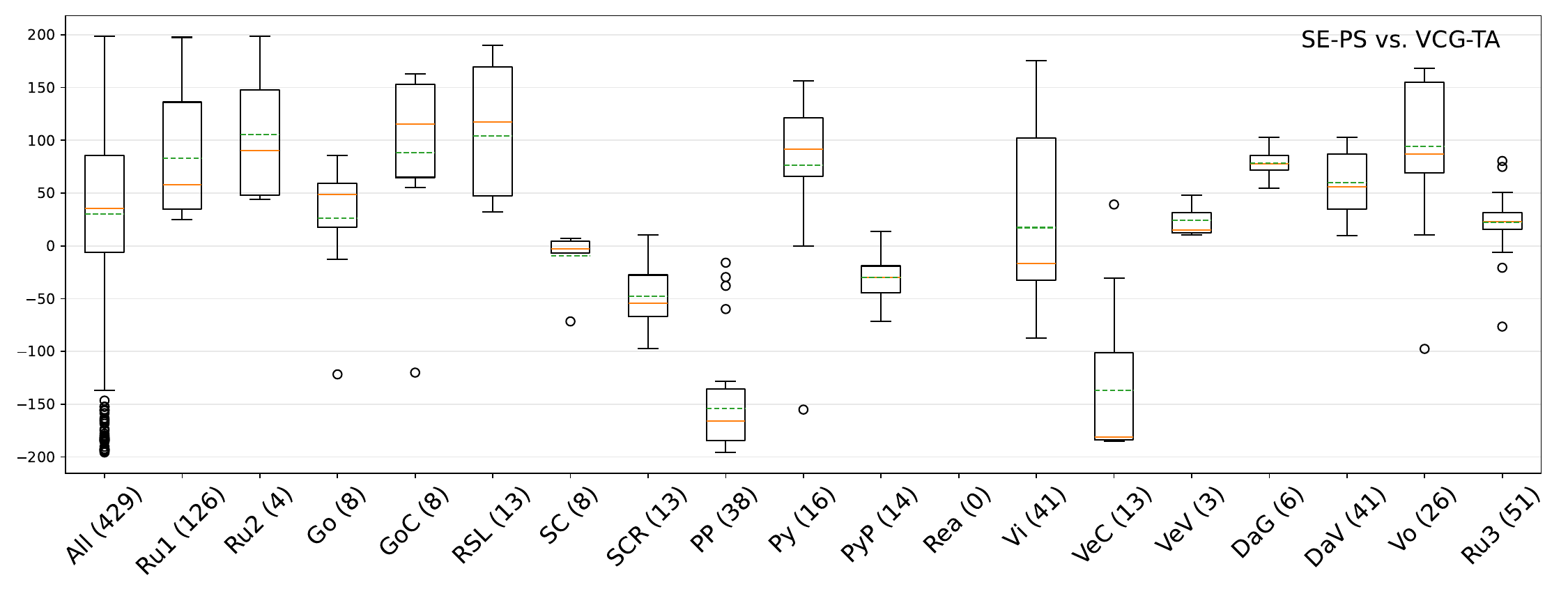}  
\caption{Box plot of relative percentage difference (RPD) of mean performance per group for the two extreme points in the design space, \greedy vs. \carbon.
Values greater than zero indicate that the first algorithm in the pair is faster, values less than zero that the second is faster;
the orange line denotes the median and the dashed green line the mean.}\label{fig:performanceExtreme} 
\end{figure}

\begin{figure}
\includegraphics[width=\textwidth]{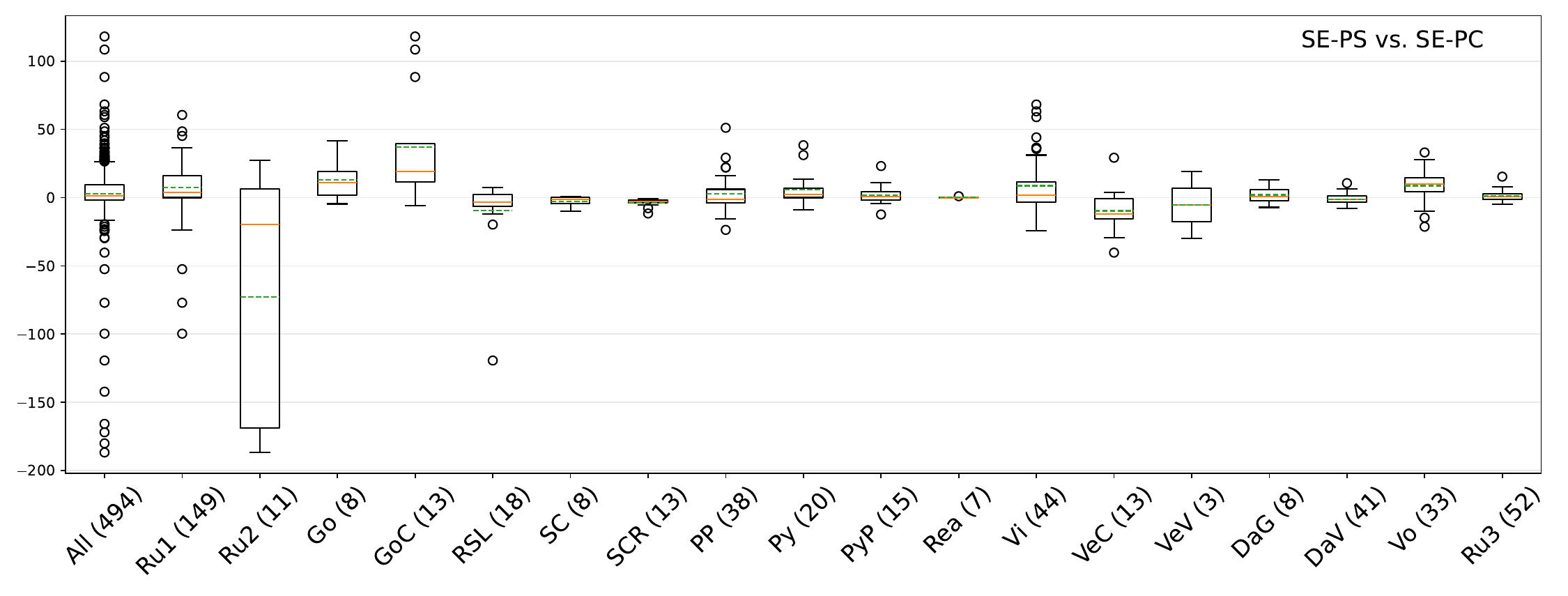}  
\includegraphics[width=\textwidth]{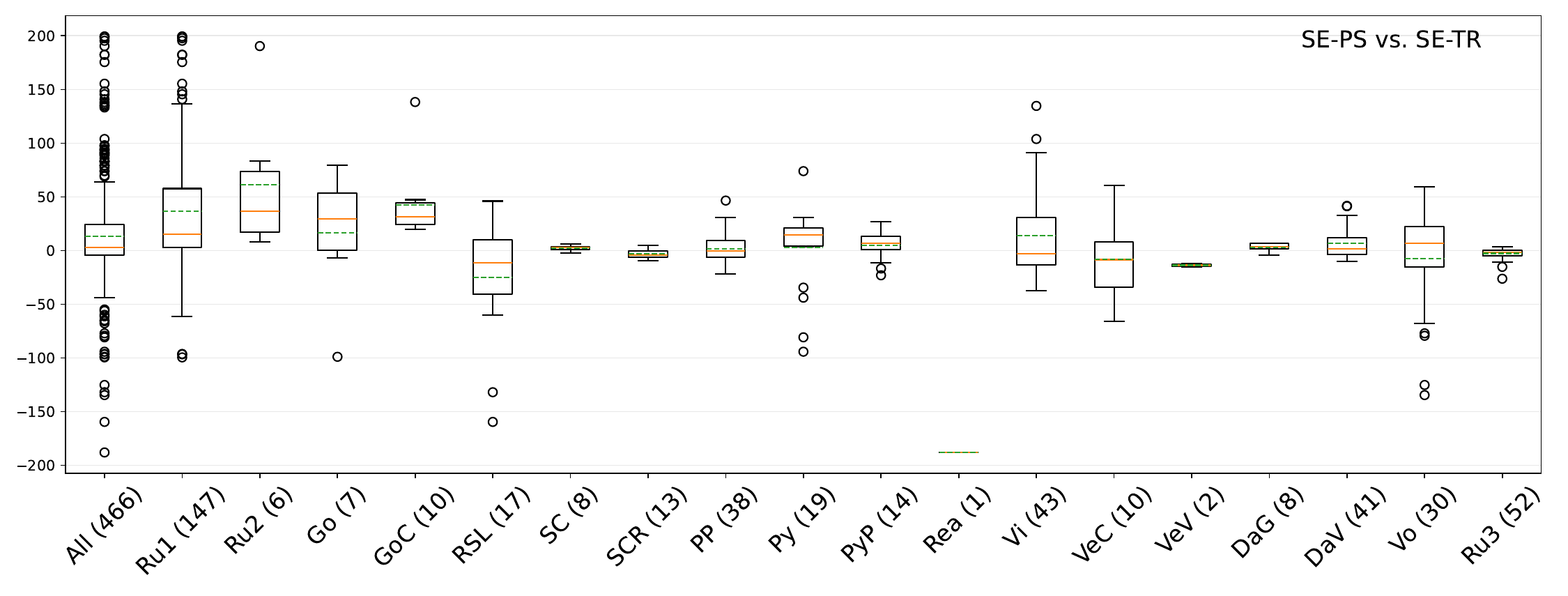}  
\includegraphics[width=\textwidth]{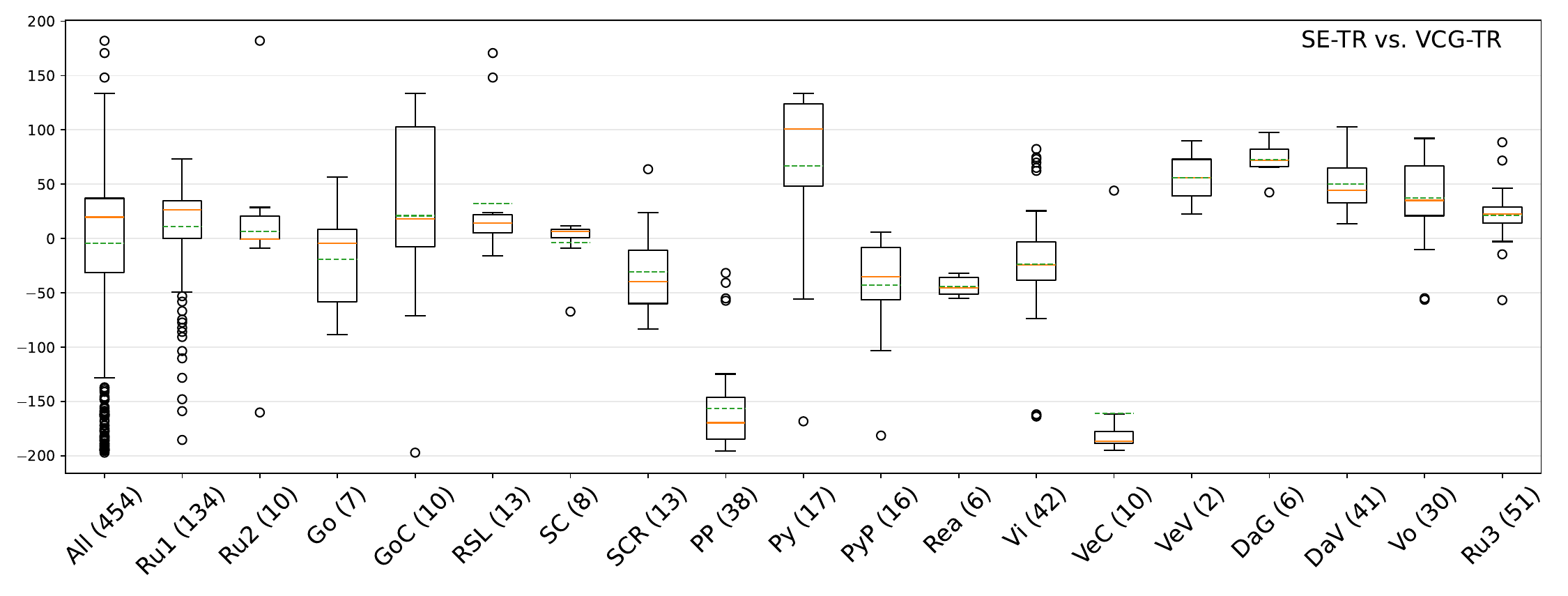}  
\includegraphics[width=\textwidth]{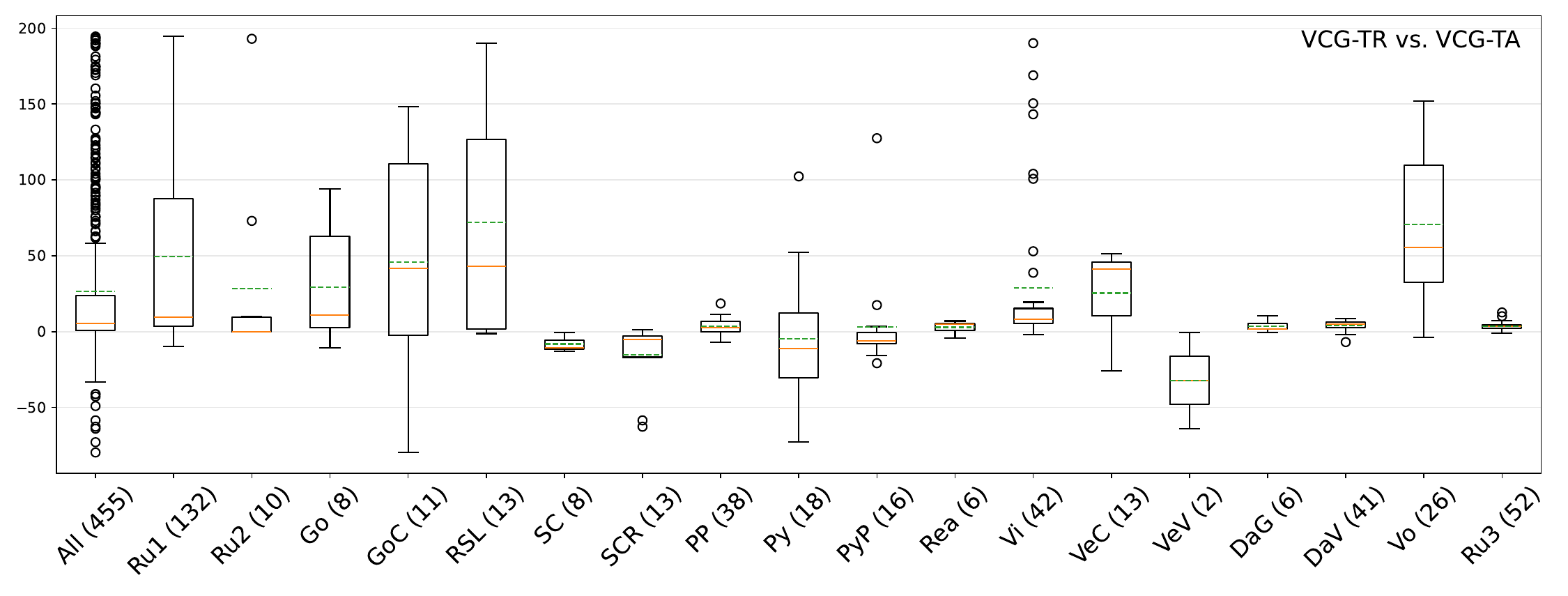}  
\caption{Box plots of relative percentage difference (RPD) of mean performance per group. From top to bottom: (a) \greedy vs.\ \mce, (b) \greedy vs.\ \sica, (c) \sica vs.\ \caco, (d) \caco vs.\ \carbon.
As before, values greater than zero indicate that the first algorithm in the pair is faster, values less than zero that the second is faster.
The orange line denotes the median, the dashed green line the mean.
}\label{fig:performanceRelated}
\end{figure}

\subsubsection{Overview.}
Fig.~\ref{fig:performanceExtreme} shows box plots of the RPDs of the two extreme points in the design space, \greedy vs.\ \carbon. Fig.~\ref{fig:performanceRelated} shows the RDPs of the closely related pairs (a) \greedy vs.\ \mce, (b) \greedy vs.\ \sica, (c) \sica vs.\ \caco, and (d) \caco vs.\ \carbon.
The orange line indicates the median, the dashed green line the mean. The lower and upper ends of the box signify the first and third quartiles ($Q_1$ and $Q_3$), respectively. 
The whiskers show the 1.5 interquartile range ($\mathit{IQR}$) values, i.e., the lowest point in the range between $Q_1 - 1.5\times \mathit{IQR}$ and $Q_1$ and the highest point between $Q_3$ and $Q_3 + 1.5\times \mathit{IQR}$, where $\mathit{IQR} = Q_3 - Q_1$.

\subsubsection{Extreme Designs.}
We first compare the two extreme points, \greedy and \carbon, which show significant performance differences across many groups (see Fig.~\ref{fig:performanceExtreme}):
\greedy performs better for most groups, in particular for \gprustiother, \gprusticp, \grsl, \gnaginiNormal, 
\gdafnygobra, \gdafnyviper, and \gvoila,
while \carbon performs significantly better for \gmpp, and to a lesser extent for \gnaginimpp, \gtwovyperRes, and \gvercorsCase.
Only \gtwovyperNoRes shows essentially the same performance for both algorithms, and \gviper contains examples that favor each of the two algorithms.
For \gmpp and \gtwovyperRes, \carbon's advantage is likely due to a high amount of branching: they have an average \malte{of} 133 and 7.5 branches per method, respectively, whereas all other groups have an average of 2 branches per method.
\gnaginimpp also results from product program constructions; the verification work again stems mostly from branching, even if the average is only 2.5 branches per method.
For \gvercorsCase, \carbon's advantage cannot be explained with branching, and we will revisit this group further below.
Thus, we conclude that \textbf{\greedy usually outperforms \carbon}, but \textbf{\carbon performs much better on branch-heavy programs}.

\subsubsection{Related Designs.}

To asses the impact of individual design decisions, we compare each algorithm to the most similar alternative(s); all comparisons can be found in Fig.~\ref{fig:performanceRelated}. 

Comparing \greedy vs.\ \mce, we observe that the two algorithms exhibit very similar performance, with \finalVersion{some small advantages} for \greedy, and a few outliers where \mce is significantly faster, in particular from \gprusticp.
This is in line with our completeness evaluation, where the two partial heap algorithms were also the most similar pair. 
Analogous to completeness, the existence of outliers again allows concluding that \textbf{being able to use more than one algorithm can result in performance advantages for individual examples}, even if the alternatives perform similar on average.

Moving from a partial heap algorithm (\greedy) to a total heap algorithm (\sica), we observe the following:
both SE algorithms exhibit similar performance for most groups;
but \greedy performs minimally better on \gprustiother, \gprusticp, and \ggobracs,
while \sica has an advantage on \grsl and \gvercors.
This leads to the surprising conclusion that the choice of \textbf{heap representation has a comparably small impact on the average performance} of an SE verifier, whereas it had a large impact on completeness.
Exceptions exist, however, and they exhibit a pattern: when \sica outperforms \greedy, it is almost exclusively on examples that heavily use \ISCs. Conversely, when \greedy is much faster, it tends to be on \gprustiother examples, which do not use \ISCs but whose heap access patterns are amenable to \greedy's optimizations.

Comparing \sica and \caco shows that \textbf{the biggest difference in average performance is caused by the switch from SE to VCG} (while keeping the same heap model). As for the two extremes (\greedy vs.\ \carbon), the branch-heavy groups (\gmpp, \gnaginimpp, and \gtwovyperRes), as well as \greach (and the aforementioned \gvercorsCase%
\footnote{We investigated this group in more detail, and observed that \finalVersion{the run time in SE algorithms} is dominated by certain SMT queries that involve heap-dependent functions and mathematical sequences, and that non-deterministically take a long time to be answered. It is ultimately unclear to us why \finalVersion{VCG} should pose a conceptual advantage here, but \finalVersion{it is plausible that for example slightly different function axiomatizations accidentally} influence how the SMT solver instances the sequence axioms, which are known to be challenging for performance.}), show a significant advantage for \caco, whereas \gnaginiNormal, \ggobracs, \gdafnygobra, \gdafnyviper and \gvoila are faster with \sica (which also slightly outperforms \caco on most of the other groups).

Finally, comparing \caco (heap per resource) to \carbon (single heap) shows very similar performance for many groups, with some exceptions (\ggobra, \ggobracs, \grsl, \gprustiother and \gvoila) that show a significant advantage for \caco \finalVersion{and no significant advantages for \carbon}. From that, we conclude that \textbf{one total heap per resource generally performs better than a single total heap}.

\subsection{Recommendation}

The overall winner of our comparison in terms of both completeness and performance is \greedy, which makes \greedy a good default algorithm, \finalVersion{followed closely by \mce}.
However, for both metrics, we have also found that
(1)~on individual examples, any of the other algorithms may outperform \greedy,
\finalVersion{(2)~some amount of optimization is required to achieve this performance and a less-optimized version of \greedy would perform worse, and, most importantly,}
(3)~there are entire categories of examples where \greedy is substantially less complete than a total-heap algorithm (\greach), or where \greedy is substantially slower than a VCG algorithm (\gmpp).
We thus recommend to either choose the algorithm based on the expected examples (\eg for domain-specific applications), or to combine \greedy with other algorithms, as discussed in Sec.~\ref{sec:portfolio}.

Our novel \sica algorithm, combining total heaps and SE, has shown that it provides a different and useful trade-off compared to existing algorithms. Its completeness is comparable to (and often slightly better than) existing total-heap algorithms (which use VCG), while
its performance is comparable to (albeit in general \finalVersion{slightly} slower than) existing SE algorithms (which use partial heaps).

Our \mce algorithm performs very similarly to \greedy, \ie very well, with some exceptions in both directions. \mce is thus also a good default algorithm, in particular, if \greedy's disjunctive aliasing incompleteness is not acceptable.

\carbon is almost universally worse than \caco and we thus recommend \caco when developing a VCG verifier.

\subsection{Threats to Validity}

\subsubsection{Benchmark Selection.}
Our evaluation covers a wide range of use cases and feature combinations, but cannot be representative of all existing (and future) examples. Our recommendation to use multiple algorithms (see also \secref{sec:portfolio}) increases the robustness of a verifier against unexpected examples.

We focused on verifying complex examples (often with quantifier-heavy specifications), whose verification time is between tenths of seconds and several minutes. As demonstrated by other tools (e.g.,  VeriFast), simpler settings (\eg without fractional permissions and \ISCs) can lead to substantially shorter verification times, in which case the differences between the algorithms might be much less pronounced. Our \gprustiother group (comparably few quantifiers, no \ISCs) comes closest to such a setting, so our results for this group should be transferable: here, the clear result is that \greedy performs the best in terms of both completeness and performance.

Verification examples are typically developed while getting feedback from the verification tool. In our case, the algorithms used by this tool were (earlier versions of) \greedy, \mce, or \carbon, since \sica and \caco were not yet implemented when any of the examples used in our evaluation were developed. This may skew the results in favor of these algorithms, because developers might have chosen designs that are handled well in the used algorithm.

\subsubsection{Impact of Optimizations and Implementation Maturity.}
\finalVersion{
The completeness and performance of the different implementations can be influenced both by optimizations they perform and by bugs they may contain.
Of the algorithms we used, the implementations of \greedy and \carbon are the most mature (in terms of development time); since these are the two best and worst performing implementations in terms of completeness, we conclude that our results are not a consequence of implementation maturity but of the algorithms themselves. It is, however, possible that the remaining three algorithms with less mature implementations could be further improved with more development time.
The fact that SE-PS and SE-PA perform the largest amount of explicit optimization is mostly because, as said before, partial-state SE-algorithms offer more potential for optimization, not because of implementation maturity; the VCG algorithms (and to a lesser degree SE-TR) leave much more work to the SMT solver, and as a result, almost no optimization beyond tuning quantifier heuristics and generating efficient VCs are possible. We have done the former for all algorithms, and Boogie is well-optimized to do the latter.
}

\subsubsection{SMT Solver.}
We have performed our evaluation using Z3 4.8.7, Viper's default solver. Different SMT solvers have different performance characteristics, but experiments with other SMT solvers determined that Z3 offers the best completeness (and performance) for our examples, and thus, was the best current choice for performing the evaluation. Future improvements in SMT solvers may disproportionately affect the evaluation results of certain algorithms: \eg improvements in quantifier reasoning may be particularly beneficial for total-heap algorithms, while improvements to incremental subsolvers may be particularly beneficial to SE algorithms.

\section{Portfolios}
\label{sec:portfolio}

The previous section showed that no single algorithm is optimal for all benchmark groups. Therefore, to maximize the chance of successful verification, it is advisable to use a \emph{portfolio} of different algorithms, i.e., to run several algorithms in parallel until at least one of them succeeds.

We explored all combinations of our five algorithms and identified four portfolios of different sizes that maximize completeness. In this section, we discuss these portfolios and evaluate their performance. It is worth noting that three out of the four winning portfolios contain \sica, which we proposed in this paper. 

Out of all possible combinations, the smallest set of algorithms needed to get the expected result for \emph{all} examples in our benchmark set is \{\greedy, \sica, \caco, \carbonx\}, which is our portfolio $P0$. Since the five algorithms we evaluated have rather diverse sources of incompleteness, most of them are needed to avoid any spurious errors in our benchmark set.

Using a large portfolio is resource intensive and not always justified in practice. There are two portfolios of size three that are complete for all but one example: \{\greedy, \sica, \cacox\} and \{\greedy, \sica, \carbonx\}. The examples they fail on seem to be very sensitive to even small changes in the verification algorithms. Each portfolio contains an SE algorithm with partial heaps, an SE algorithm with total heaps, and a VCG algorithm, which demonstrates the complementarity of these approaches. We select the first of these portfolios due to the better average performance of \caco over \carbon and name it $P1$.

Reducing the portfolio size further, we can identify two interesting portfolios of size two. The best portfolio of size two, \{\greedy, \cacox\} ($P2$), is complete for all but six examples. \{\greedy, \sicax\} ($P3$) is complete for all but ten examples, but has the major advantage that it contains only SE algorithms, which reduces the effort of implementing this portfolio substantially. Different SE algorithms can share many parts of the implementation, whereas the implementations of SE and VCG algorithms offer little opportunity for reuse. Consequently, $P3$ provides a relatively easy way for SE tools to improve their completeness by complementing their existing \greedy algorithm with the new \sica algorithm, which only requires re-implementing the heap operations.

\ifarxiv App.~\ref{app:portfolioPerf} \fi shows the relative performance of $P1$, $P2$, and $P3$ versus $P0$.
While $P1$ mostly performs identical to $P0$, $P2$ has a slight disadvantage for \grsl and \gvercorsCase, but performs equally well for all other groups, and thus delivers almost all the benefits of $P0$. Finally, $P3$, due to its lack of a VCG verifier, performs much worse than $P0$ in the previously-identified branch-heavy groups, while delivering good performance for all others.

\section{Related Work}
\label{sec:related}

The first verification algorithm for separation logic was the SE algorithm for Smallfoot~\cite{BerdineCO05}. Its partial heap representation as a collection of heap chunks has been adopted and refined by many separation logic verifiers (such as  Caper~\cite{Dinsdale-YoungP17}, the Gillian instantiations for C and JavaScript~\cite{MaksimovicASG20}, JaVerT~\cite{SantosMSG19}, SecC~\cite{ErnstM19}, VeriFast~\cite{JacobsSPVPP11}, and Viper's SE-backend), for instance, to support user-defined predicates~\cite{JacobsSPVPP11}, alternative permission models such as fractional permissions~\cite{Boyland03} and counting permissions~\cite{BornatCOP05}, advanced separation logic connectives such as magic wands~\cite{wands} and iterated separating conjunction~\cite{0001SS16}, and proof search for angelic choice using backtracking~\cite{Dinsdale-YoungP17}. Our evaluation covers those extensions that are implemented in Viper, namely predicates, fractional permissions, magic wands, and iterated separating conjunction.

Gillian~\cite{SantosMAG20} is an SE framework 
that can be instantiated for different input languages and separation logics %
and lets 
each instantiation define its own representation of the heap.
The existing instantiations for C and JavaScript use an algorithm \lastMinute{similar but not identical to \greedy}, but Gillian could also express the other SE algorithms we discuss. 
Our evaluation can guide developers toward an optimal use of Gillian's expressiveness.

VCG for separation logic was first developed in the context of VeriCool~\cite{SmansJP09} and then extended to concurrency in Chalice~\cite{LeinoMS09}. 
The algorithm there, as well as in Viper's VCG-backend, uses a total-heap representation. %
While Chalice and Viper use a single total map to represent all heap values, GrassHopper~\cite{PiskacWZ14} uses a dedicated map for each resource. As we observed in our evaluation, and has previously been shown in the context of VCC~\cite{BohmeM11}, this representation can improve performance and completeness by simplifying framing. GrassHopper uses advanced algorithms to automate reasoning about predicates, which were not in scope for our evaluation here.

Existing verifiers support a range of permission logics, including separation logic and implicit dynamic frames~\cite{SmansJP09}.
Separation logic is typically defined over partial heaps, whereas the theory of implicit dynamic frames uses total heaps. However, there is a strong connection between both logics~\cite{ParkinsonS11}, and the algorithms discussed in this paper can support both.

There are other approaches to automating verification in separation logic. For instance, Steel~\cite{FromherzRSGMMR21} is built on top of F*~\cite{SwamyWSCL13}, which uses type inference to devise derivations in a dependently typed separation logic. 
RefinedC~\cite{SammlerLKMD021} automates proof search in Lithium, a fragment of the Iris separation logic~\cite{JungKJBBD18}. Its verification algorithm is implemented in Coq as a tactic.
Such approaches differ substantially from the SE and VCG algorithms discussed in this paper in the degree of automation they provide, their expressiveness, or their ability to devise foundational proofs, which makes a meaningful comparison difficult.
\finalVersion{
Hip/Sleek~\cite{DBLP:journals/scp/ChinDNQ12} performs a forward verification similar to SE but operates directly by checking entailments on separation logic formulas and, thus, does not need a heap encoding.
To our knowledge, existing separation logic solvers do not support all of the separation logic features (predicates, magic wands and quantified resources, and their combination) supported by the algorithms we considered.}

Kassios et al.~\cite{KassiosMS12} compared the performance of Chalice's VCG algorithm to an alternative SE-backend most similar to \greedy, and found a significant performance advantage for the SE-backend throughout. However, their comparison does not include the versions of SE and VCG used in modern tools, does not assess completeness, and does not reflect the diversity of verification problems, with only 29 examples in total being compared.

\finalVersion{
Finally, choices between explicitly enumerating states (\eg heap chunks in partial heap models and program paths in SE) and using logical formulas to represent the different options (in total heap models and VCG algorithms) also exist for other ways of automated reasoning. 
For example, explicit-state model checking enumerates individual states, whereas symbolic model checking represents sets of states via logical formulas, offering different tradeoffs~\cite{DBLP:conf/iecon/BuzhinskyPV17}.
}

\section{Conclusions and Future Work}
\label{sec:conclusions}

We have presented and implemented five algorithms for automated separation logic verification, including two novel algorithms. Our evaluation shows that, across all benchmarks, the prevalent \greedy algorithm shows the best completeness and performance. However, it is not optimal for all benchmark groups and, thus, should be complemented by other algorithms. We identified algorithm portfolios of different sizes that maximize completeness.

As future work, we plan to extract features from programs that allow us to predict which algorithm will perform best. 

\subsubsection{\ackname}

We are grateful to Sacha-Elie Ayoun, Thomas Dinsdale-Young, and Thomas Wies for discussions about Gillian, Caper, and GrassHopper.
We thank Robin Sierra for a first implementation of \mce.
\finalVersion{We thank the ETH Seminar for Statistics consulting service for helpful discussions.}

\newpage
\bibliographystyle{splncs04}
\bibliography{references}

\begin{thebibliography}{10}
\providecommand{\url}[1]{\texttt{#1}}
\providecommand{\urlprefix}{URL }
\providecommand{\doi}[1]{https://doi.org/#1}

\bibitem{vercors-redblack}
Armborst, L., Huisman, M.: Permission-based verification of red-black trees and
  their merging. In: FormaliSE@ICSE. pp. 111--123. {IEEE} (2021)

\bibitem{wireguard}
Arquint, L., Schwerhoff, M., Mehta, V., M{\"{u}}ller, P.: A generic methodology
  for the modular verification of security protocol implementations. In: {CCS}.
  pp. 1377--1391. {ACM} (2023)

\bibitem{vytautas-thesis}
Astrauskas, V.: Leveraging Uniqueness for Modular Verification of
  Heap-Manipulating Programs. Ph.D. thesis, {ETH} Zurich, Z{\"{u}}rich,
  Switzerland (2024)

\bibitem{prusti}
Astrauskas, V., M{\"{u}}ller, P., Poli, F., Summers, A.J.: Leveraging {R}ust
  types for modular specification and verification. Proc. {ACM} Program. Lang.
  \textbf{3}({OOPSLA}),  147:1--147:30 (2019)

\bibitem{CVC5}
Barbosa, H., Barrett, C.W., Brain, M., Kremer, G., Lachnitt, H., Mann, M.,
  Mohamed, A., Mohamed, M., Niemetz, A., N{\"{o}}tzli, A., Ozdemir, A.,
  Preiner, M., Reynolds, A., Sheng, Y., Tinelli, C., Zohar, Y.: {CVC5}: {A}
  versatile and industrial-strength {SMT} solver. In: {TACAS} {(1)}. Lecture
  Notes in Computer Science, vol. 13243, pp. 415--442. Springer (2022)

\bibitem{boogie}
Barnett, M., Chang, B.E., DeLine, R., Jacobs, B., Leino, K.R.M.: Boogie: {A}
  modular reusable verifier for object-oriented programs. In: {FMCO}. LNCS,
  vol.~4111, pp. 364--387. Springer (2005)

\bibitem{BarnettL05}
Barnett, M., Leino, K.R.M.: Weakest-precondition of unstructured programs. In:
  {PASTE}. pp. 82--87. {ACM} (2005)

\bibitem{BerdineCO05}
Berdine, J., Calcagno, C., O'Hearn, P.W.: Smallfoot: Modular automatic
  assertion checking with separation logic. In: {FMCO}. Lecture Notes in
  Computer Science, vol.~4111, pp. 115--137. Springer (2005)

\bibitem{vercors}
Blom, S., Huisman, M.: The {VerCors} tool for verification of concurrent
  programs. In: {FM}. Lecture Notes in Computer Science, vol.~8442, pp.
  127--131. Springer (2014)

\bibitem{BohmeM11}
B{\"{o}}hme, S., Moskal, M.: Heaps and data structures: {A} challenge for
  automated provers. In: {CADE}. Lecture Notes in Computer Science, vol.~6803,
  pp. 177--191. Springer (2011)

\bibitem{BornatCOP05}
Bornat, R., Calcagno, C., O'Hearn, P.W., Parkinson, M.J.: Permission accounting
  in separation logic. In: {POPL}. pp. 259--270. {ACM} (2005)

\bibitem{Boyland03}
Boyland, J.: Checking interference with fractional permissions. In: {SAS}.
  Lecture Notes in Computer Science, vol.~2694, pp. 55--72. Springer (2003)

\bibitem{2vyper}
Br{\"{a}}m, C., Eilers, M., M{\"{u}}ller, P., Sierra, R., Summers, A.J.: Rich
  specifications for {E}thereum smart contract verification. Proc. {ACM}
  Program. Lang.  \textbf{5}({OOPSLA}),  1--30 (2021)

\bibitem{DBLP:conf/iecon/BuzhinskyPV17}
Buzhinsky, I., Pakonen, A., Vyatkin, V.: Explicit-state and symbolic model
  checking of nuclear i{\&}c systems: {A} comparison. In: {IECON}. pp.
  5439--5446. {IEEE} (2017)

\bibitem{DBLP:journals/scp/ChinDNQ12}
Chin, W., David, C., Nguyen, H.H., Qin, S.: Automated verification of shape,
  size and bag properties via user-defined predicates in separation logic. Sci.
  Comput. Program.  \textbf{77}(9),  1006--1036 (2012)

\bibitem{ematching}
Detlefs, D., Nelson, G., Saxe, J.B.: Simplify: a theorem prover for program
  checking. J. {ACM}  \textbf{52}(3),  365--473 (2005)

\bibitem{Dinsdale-YoungP17}
Dinsdale{-}Young, T., da~Rocha~Pinto, P., Andersen, K.J., Birkedal, L.: Caper -
  automatic verification for fine-grained concurrency. In: {ESOP}. Lecture
  Notes in Computer Science, vol. 10201, pp. 420--447. Springer (2017)

\bibitem{verifythis}
Dross, C., Furia, C.A., Huisman, M., Monahan, R., M{\"{u}}ller, P.:
  {V}erify{T}his 2019: a program verification competition. Int. J. Softw. Tools
  Technol. Transf.  \textbf{23}(6),  883--893 (2021)

\bibitem{egli}
Egli, T.: Translating Pedagogical Exercises to Viper's Go Front-End. Bachelor's
  thesis, ETH Z{\"u}rich (2023)

\bibitem{marco-thesis}
Eilers, M.: Modular Specification and Verification of Security Properties for
  Mainstream Languages. Ph.D. thesis, {ETH} Zurich, Z{\"{u}}rich, Switzerland
  (2022)

\bibitem{nagini-mpp}
Eilers, M., Meier, S., M{\"{u}}ller, P.: Product programs in the wild:
  Retrofitting program verifiers to check information flow security. In: {CAV}
  {(1)}. Lecture Notes in Computer Science, vol. 12759, pp. 718--741. Springer
  (2021)

\bibitem{nagini}
Eilers, M., M{\"{u}}ller, P.: Nagini: {A} static verifier for {P}ython. In:
  {CAV} {(1)}. Lecture Notes in Computer Science, vol. 10981, pp. 596--603.
  Springer (2018)

\bibitem{mpp}
Eilers, M., M{\"{u}}ller, P., Hitz, S.: Modular product programs. {ACM} Trans.
  Program. Lang. Syst.  \textbf{42}(1),  3:1--3:37 (2020)

\bibitem{ErnstM19}
Ernst, G., Murray, T.: {S}ec{CSL}: Security concurrent separation logic. In:
  {CAV} {(2)}. Lecture Notes in Computer Science, vol. 11562, pp. 208--230.
  Springer (2019)

\bibitem{FilliatreP13}
Filli{\^{a}}tre, J., Paskevich, A.: Why3 - where programs meet provers. In:
  {ESOP}. Lecture Notes in Computer Science, vol.~7792, pp. 125--128. Springer
  (2013)

\bibitem{forster}
Forster, S.: Static Verification of the {SCION} Router Implementation.
  Bachelor's thesis, ETH Z{\"u}rich (2018)

\bibitem{frei}
Frei, B.: Translating Pedagogical Verification Exercises to Viper. Bachelor's
  thesis, ETH Z{\"u}rich (2023)

\bibitem{FromherzRSGMMR21}
Fromherz, A., Rastogi, A., Swamy, N., Gibson, S., Mart{\'{\i}}nez, G.,
  Merigoux, D., Ramananandro, T.: Steel: proof-oriented programming in a
  dependently typed concurrent separation logic. Proc. {ACM} Program. Lang.
  \textbf{5}({ICFP}),  1--30 (2021)

\bibitem{HeuleKMS13}
Heule, S., Kassios, I.T., M{\"{u}}ller, P., Summers, A.J.: Verification
  condition generation for permission logics with abstract predicates and
  abstraction functions. In: {ECOOP}. Lecture Notes in Computer Science,
  vol.~7920, pp. 451--476. Springer (2013)

\bibitem{JacobsSPVPP11}
Jacobs, B., Smans, J., Philippaerts, P., Vogels, F., Penninckx, W., Piessens,
  F.: {V}eri{F}ast: {A} powerful, sound, predictable, fast verifier for {C} and
  {J}ava. In: {NASA} Formal Methods. Lecture Notes in Computer Science,
  vol.~6617, pp. 41--55. Springer (2011)

\bibitem{JungKJBBD18}
Jung, R., Krebbers, R., Jourdan, J., Bizjak, A., Birkedal, L., Dreyer, D.: Iris
  from the ground up: {A} modular foundation for higher-order concurrent
  separation logic. J. Funct. Program.  \textbf{28}, ~e20 (2018)

\bibitem{KassiosMS12}
Kassios, I.T., M{\"{u}}ller, P., Schwerhoff, M.: Comparing verification
  condition generation with symbolic execution: An experience report. In:
  {VSTTE}. Lecture Notes in Computer Science, vol.~7152, pp. 196--208. Springer
  (2012)

\bibitem{LalQ14}
Lal, A., Qadeer, S.: Powering the static driver verifier using {C}orral. In:
  {SIGSOFT} {FSE}. pp. 202--212. {ACM} (2014)

\bibitem{LeinoBoogie2}
Leino, K.R.M.: This is {B}oogie 2 (June 2008),
  \url{https://www.microsoft.com/en-us/research/publication/this-is-boogie-2-2/}

\bibitem{Leino10}
Leino, K.R.M.: Dafny: An automatic program verifier for functional correctness.
  In: {LPAR} (Dakar). Lecture Notes in Computer Science, vol.~6355, pp.
  348--370. Springer (2010)

\bibitem{leino2023program}
Leino, K.R.M.: Program Proofs. MIT Press (2023)

\bibitem{LeinoMS09}
Leino, K.R.M., M{\"{u}}ller, P., Smans, J.: Verification of concurrent programs
  with {C}halice. In: {FOSAD}. Lecture Notes in Computer Science, vol.~5705,
  pp. 195--222. Springer (2009)

\bibitem{MaksimovicASG20}
Maksimovic, P., Ayoun, S., Santos, J.F., Gardner, P.: Gillian, part {II:}
  real-world verification for {J}ava{S}cript and {C}. In: {CAV} {(2)}. Lecture
  Notes in Computer Science, vol. 12760, pp. 827--850. Springer (2021)

\bibitem{Z3}
de~Moura, L.M., Bj{\o}rner, N.S.: {Z3:} an efficient {SMT} solver. In: {TACAS}.
  Lecture Notes in Computer Science, vol.~4963, pp. 337--340. Springer (2008)

\bibitem{portfolio24}
Mugnier, E., McLaughlin, S., Tomb, A.: Portfolio solving for {D}afny. In: Dafny
  Workshop (2024), to appear

\bibitem{0001SS16}
M{\"{u}}ller, P., Schwerhoff, M., Summers, A.J.: Automatic verification of
  iterated separating conjunctions using symbolic execution. In: {CAV} {(1)}.
  Lecture Notes in Computer Science, vol.~9779, pp. 405--425. Springer (2016)

\bibitem{viper}
M{\"{u}}ller, P., Schwerhoff, M., Summers, A.J.: Viper: {A} verification
  infrastructure for permission-based reasoning. In: {VMCAI}. Lecture Notes in
  Computer Science, vol.~9583, pp. 41--62. Springer (2016)

\bibitem{ParkinsonB05}
Parkinson, M.J., Bierman, G.M.: Separation logic and abstraction. In: {POPL}.
  pp. 247--258. {ACM} (2005)

\bibitem{ParkinsonS11}
Parkinson, M.J., Summers, A.J.: The relationship between separation logic and
  implicit dynamic frames. In: {ESOP}. Lecture Notes in Computer Science,
  vol.~6602, pp. 439--458. Springer (2011)

\bibitem{verifiedscion}
Pereira, J.C., Klenze, T., Giampietro, S., Limbeck, M., Spiliopoulos, D., Wolf,
  F.A., Eilers, M., Sprenger, C., Basin, D., Müller, P., Perrig, A.: Protocols
  to code: Formal verification of a next-generation internet router (2024)

\bibitem{PiskacWZ14}
Piskac, R., Wies, T., Zufferey, D.: Grasshopper - complete heap verification
  with mixed specifications. In: {TACAS}. Lecture Notes in Computer Science,
  vol.~8413, pp. 124--139. Springer (2014)

\bibitem{mendel-paper}
Poli, F., Denis, X., Müller, P., Summers, A.J.: Reasoning about interior
  mutability in {R}ust using library-defined capabilities (2024)

\bibitem{Reynolds02a}
Reynolds, J.C.: Separation logic: {A} logic for shared mutable data structures.
  In: {LICS}. pp. 55--74. {IEEE} Computer Society (2002)

\bibitem{tada}
da~Rocha~Pinto, P., Dinsdale{-}Young, T., Gardner, P.: {TaDA}: {A} logic for
  time and data abstraction. In: {ECOOP}. Lecture Notes in Computer Science,
  vol.~8586, pp. 207--231. Springer (2014)

\bibitem{vercors-permutation}
Safari, M., Huisman, M.: A generic approach to the verification of the
  permutation property of sequential and parallel swap-based sorting
  algorithms. In: {IFM}. Lecture Notes in Computer Science, vol. 12546, pp.
  257--275. Springer (2020)

\bibitem{vercors-prefixsum}
Safari, M., Huisman, M.: Formal verification of parallel prefix sum and stream
  compaction algorithms in {CUDA}. Theor. Comput. Sci.  \textbf{912},  81--98
  (2022)

\bibitem{SammlerLKMD021}
Sammler, M., Lepigre, R., Krebbers, R., Memarian, K., Dreyer, D., Garg, D.:
  {R}efined{C}: automating the foundational verification of {C} code with
  refined ownership types. In: {PLDI}. pp. 158--174. {ACM} (2021)

\bibitem{SantosMAG20}
Santos, J.F., Maksimovic, P., Ayoun, S., Gardner, P.: Gillian, part i: a
  multi-language platform for symbolic execution. In: {PLDI}. pp. 927--942.
  {ACM} (2020)

\bibitem{SantosMSG19}
Santos, J.F., Maksimovic, P., Sampaio, G., Gardner, P.: {J}a{V}er{T} 2.0:
  compositional symbolic execution for {J}ava{S}cript. Proc. {ACM} Program.
  Lang.  \textbf{3}({POPL}),  66:1--66:31 (2019)

\bibitem{Schwerhoff16}
Schwerhoff, M.: Advancing Automated, Permission-Based Program Verification
  Using Symbolic Execution. Ph.D. thesis, {ETH} Zurich, Z{\"{u}}rich,
  Switzerland (2016)

\bibitem{wands}
Schwerhoff, M., Summers, A.J.: Lightweight support for magic wands in an
  automatic verifier. In: {ECOOP}. LIPIcs, vol.~37, pp. 614--638. Schloss
  Dagstuhl - Leibniz-Zentrum f{\"{u}}r Informatik (2015)

\bibitem{SmansJP09}
Smans, J., Jacobs, B., Piessens, F.: Implicit dynamic frames: Combining dynamic
  frames and separation logic. In: {ECOOP}. Lecture Notes in Computer Science,
  vol.~5653, pp. 148--172. Springer (2009)

\bibitem{igloo}
Sprenger, C., Klenze, T., Eilers, M., Wolf, F.A., M{\"{u}}ller, P., Clochard,
  M., Basin, D.A.: Igloo: soundly linking compositional refinement and
  separation logic for distributed system verification. Proc. {ACM} Program.
  Lang.  \textbf{4}({OOPSLA}),  152:1--152:31 (2020)

\bibitem{rsl2viper}
Summers, A.J., M{\"{u}}ller, P.: Automating deductive verification for
  weak-memory programs. In: {TACAS} {(1)}. Lecture Notes in Computer Science,
  vol. 10805, pp. 190--209. Springer (2018)

\bibitem{SwamyWSCL13}
Swamy, N., Weinberger, J., Schlesinger, C., Chen, J., Livshits, B.: Verifying
  higher-order programs with the {D}ijkstra monad. In: {PLDI}. pp. 387--398.
  {ACM} (2013)

\bibitem{reachability}
Ter{-}Gabrielyan, A., Summers, A.J., M{\"{u}}ller, P.: Modular verification of
  heap reachability properties in separation logic. Proc. {ACM} Program. Lang.
  \textbf{3}({OOPSLA}),  121:1--121:28 (2019)

\bibitem{RSL}
Vafeiadis, V., Narayan, C.: Relaxed separation logic: a program logic for {C11}
  concurrency. In: {OOPSLA}. pp. 867--884. {ACM} (2013)

\bibitem{gobra}
Wolf, F.A., Arquint, L., Clochard, M., Oortwijn, W., Pereira, J.C.,
  M{\"{u}}ller, P.: Gobra: Modular specification and verification of {G}o
  programs. In: {CAV} {(1)}. Lecture Notes in Computer Science, vol. 12759, pp.
  367--379. Springer (2021)

\bibitem{voila}
Wolf, F.A., Schwerhoff, M., M{\"{u}}ller, P.: Concise outlines for a complex
  logic: {A} proof outline checker for {TaDA}  \textbf{13047},  407--426 (2021)

\end{thebibliography}

\ifarxiv 
\appendix

\section{Portfolio Performance Comparison}\label{app:portfolioPerf}

\figref{fig:portfolioPerformance} compares the performance of $P0$, the smallest portfolio that is complete for all examples, to those of the smaller portfolios $P1$ (of size 3) and $P2$ and $P3$ (both of size 2).
As discussed before, $P1$ and (with slightly bigger discrepancies) $P2$ show similar performance to $P0$ for most groups. Between $P0$ and $P3$, there is a substantial disadvantage for $P3$ for branch-heavy groups due to $P3$'s exclusive reliance on SE over VCG.

\figref{fig:greedyVsPortfolio} additionally compares the performance of \greedy, the best individual algorithm, to the portfolios of sizes 3 and 2. 
The comparisons with $P1$ and $P2$ show that both portfolios, in addition to being more complete, offer significant performance advantages compared to even the best individual algorithm (in particular in the aforementioned branch-heavy groups). $P3$, as expected, shows similar performance to \greedy except for some outliers (while still offering a significant completeness advantage). 

\begin{figure}
\includegraphics[width=\textwidth]{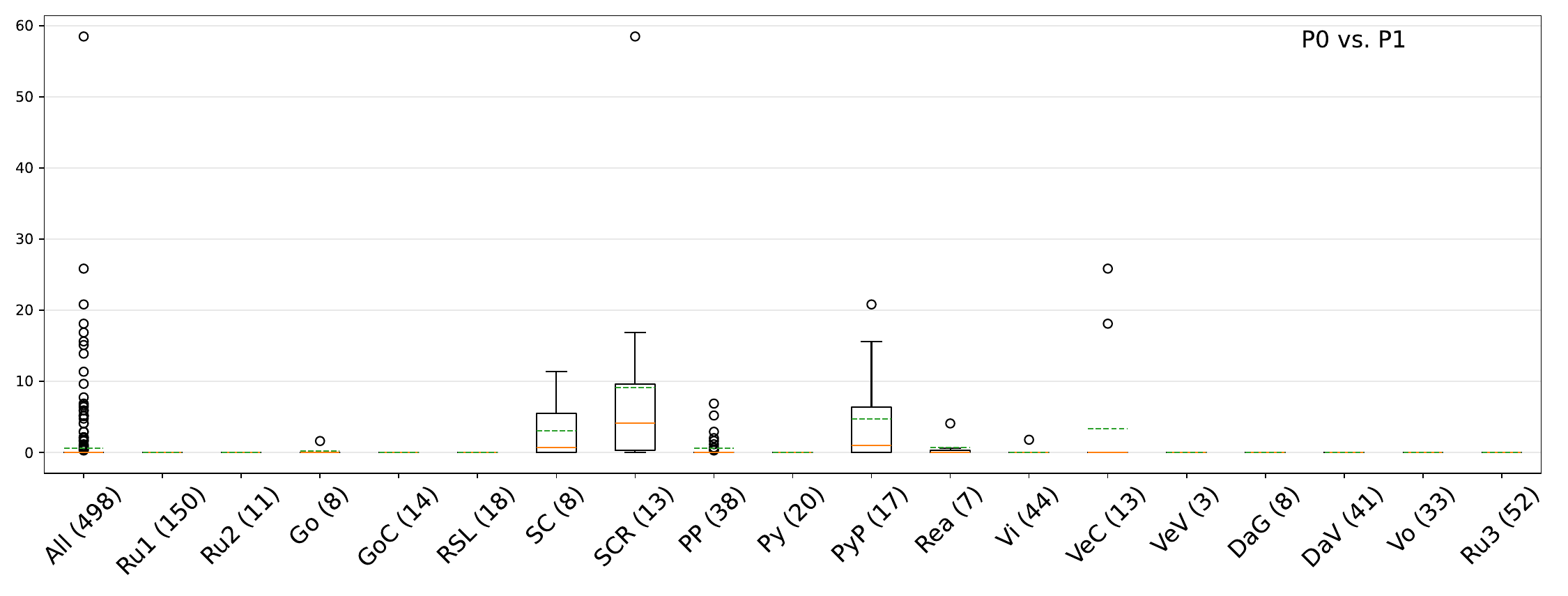}  
\includegraphics[width=\textwidth]{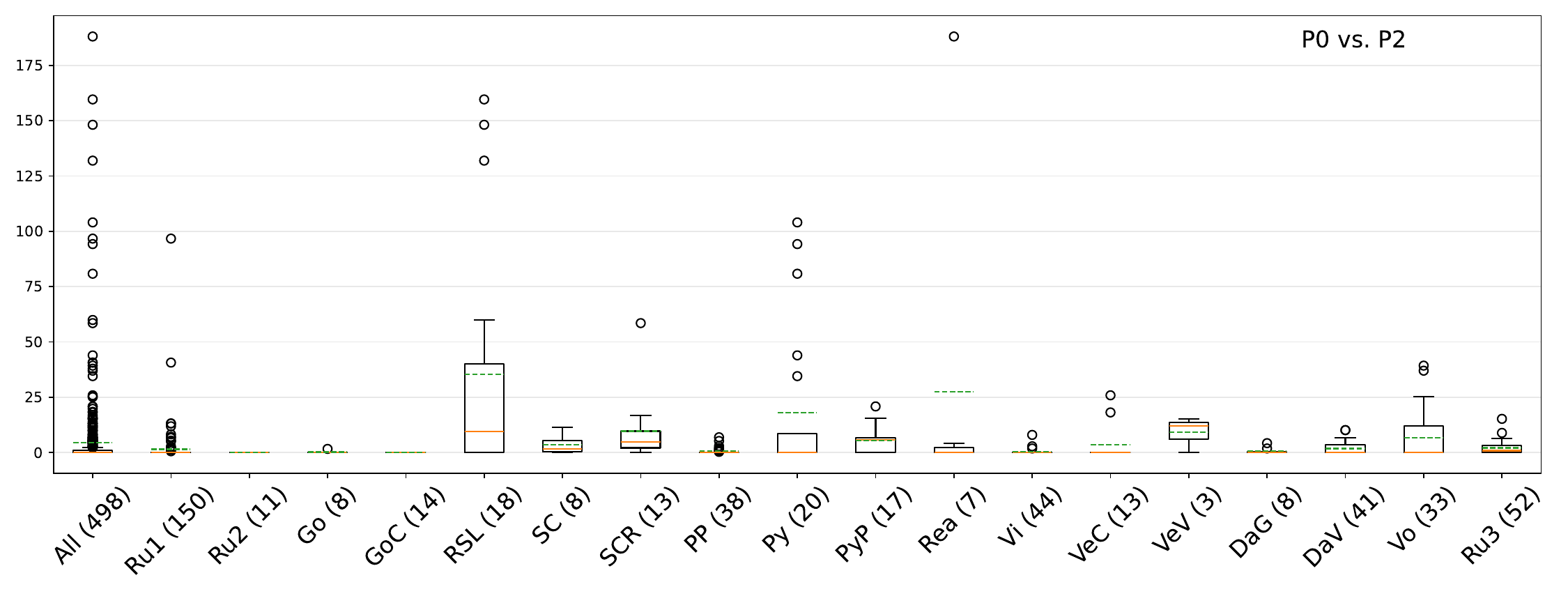}  
\includegraphics[width=\textwidth]{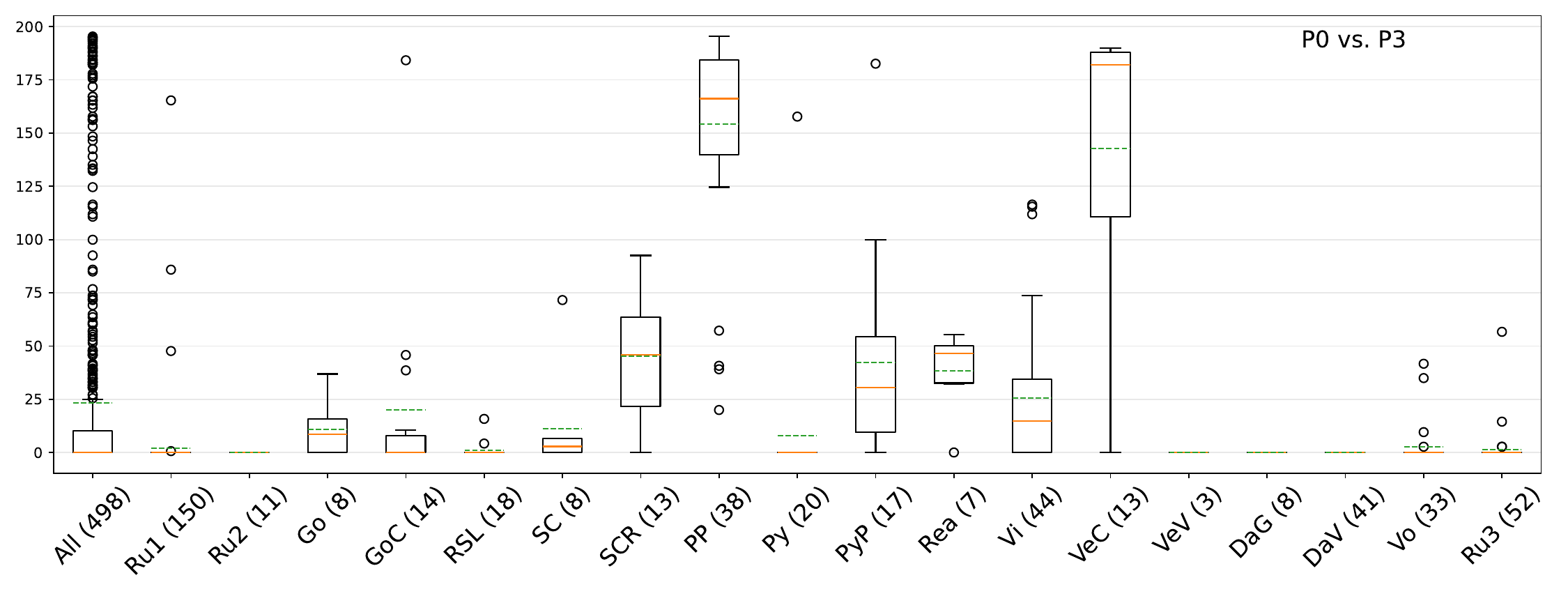}  
\caption{Box plots of relative percentage difference (RPD) of mean performance for pairs of portfolios. From top to bottom: a) $P0$ vs. $P1$, b) $P0$ vs. $P2$, c) $P0$ vs. $P3$.}\label{fig:portfolioPerformance}
\end{figure}

\begin{figure}
\includegraphics[width=\textwidth]{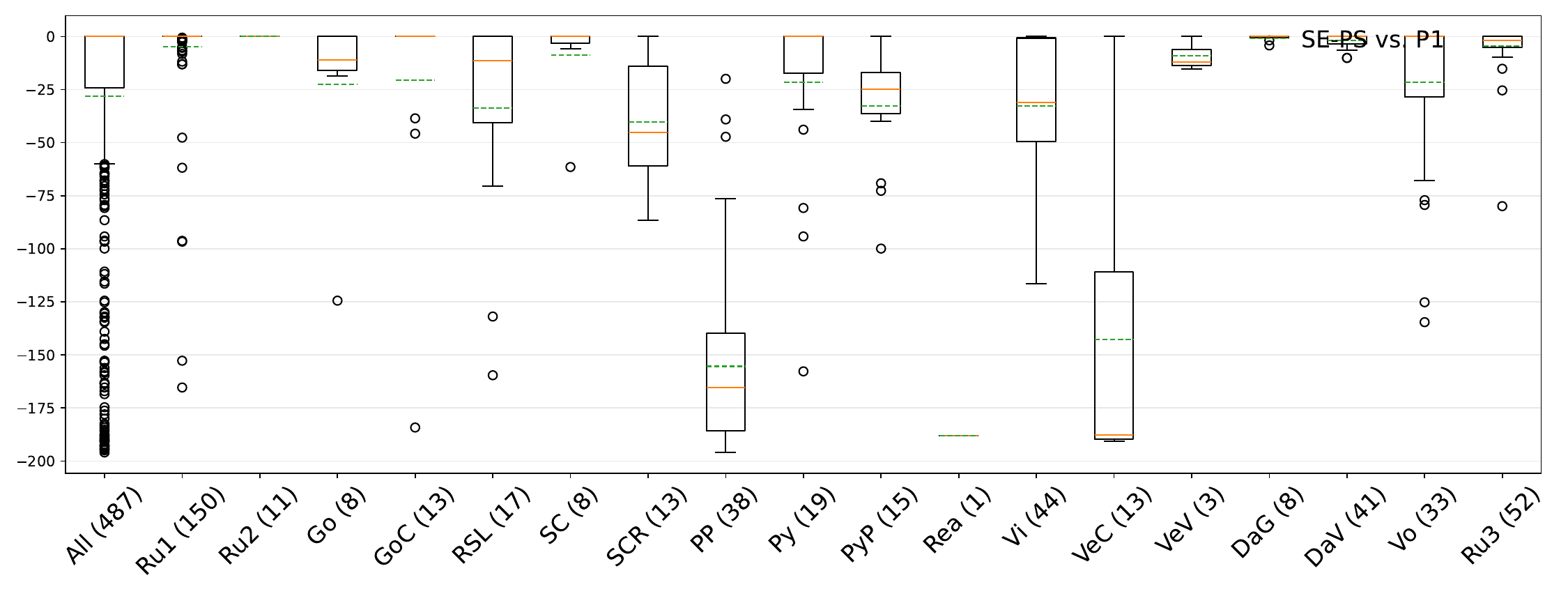}  
\includegraphics[width=\textwidth]{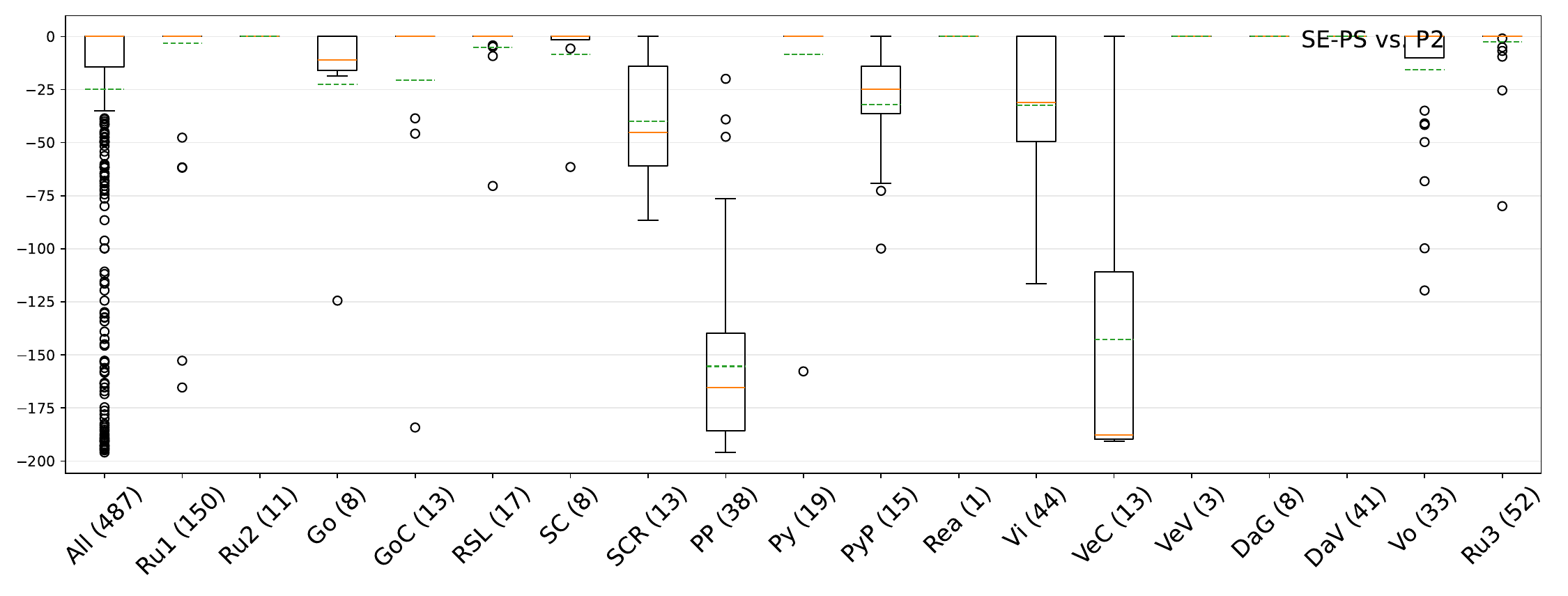}  
\includegraphics[width=\textwidth]{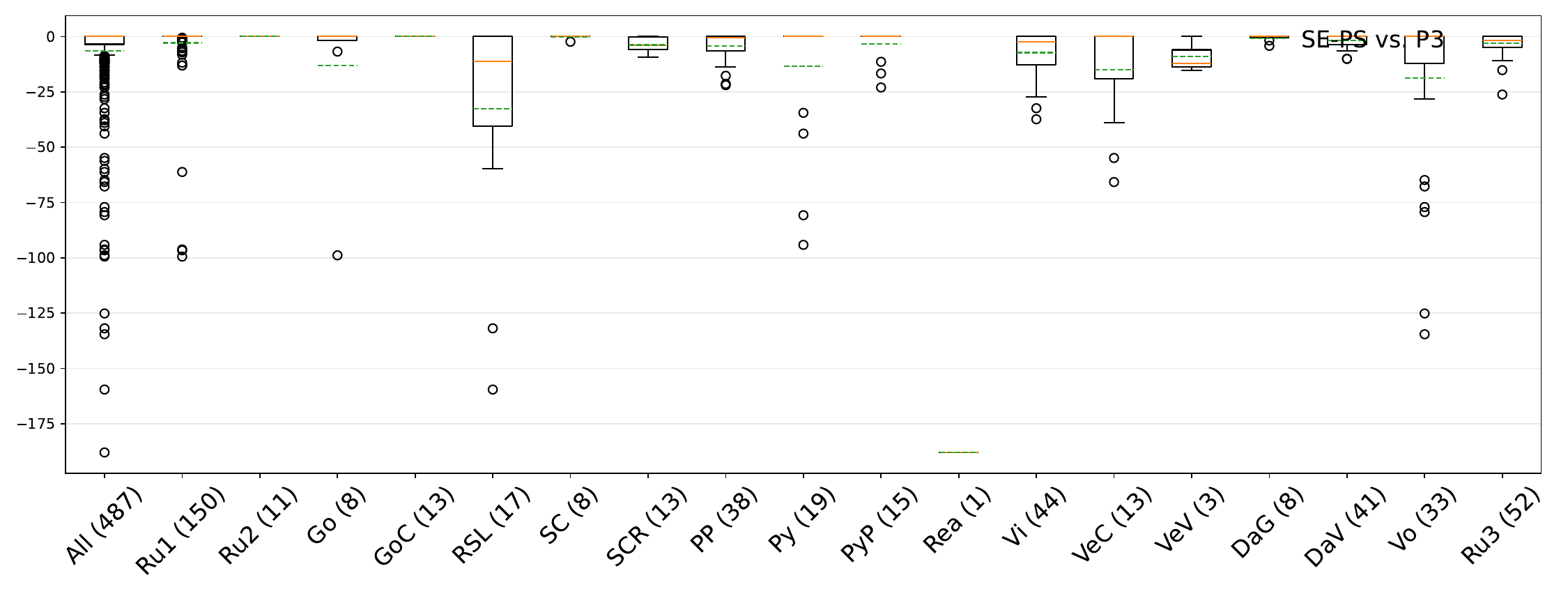}  
\caption{Box plots of relative percentage difference (RPD) of mean performance comparing \greedy to different portfolios. From top to bottom: a) \greedy vs. $P1$, b) \greedy vs. $P2$, c) \greedy vs. $P3$.}\label{fig:greedyVsPortfolio}
\end{figure}

\else \fi
\end{document}